\documentclass[preprint,10pt,5p,twocolumn]{elsarticle}

\usepackage{amssymb}
\usepackage{epsfig}
\usepackage{lineno}
\usepackage{hyperref}
\hypersetup{
    colorlinks=true,
    linkcolor=blue,
    filecolor=magenta,      
    urlcolor=cyan,
}

\journal{Journal of Solid State Chemistry}

\graphicspath{{ }}

\begin{document}

\begin{frontmatter}


\title{The intercalation chemistry of layered iron chalcogenide superconductors}

\author{Hector K. Vivanco}
\author{Efrain E. Rodriguez\corref{cor}}
\ead{efrain@umd.edu}

\address{Department of Chemistry \& Biochemistry, University of Maryland, College Park, Maryland 20742, United States}

\begin{abstract}

The iron chalcogenides FeSe and FeS are superconductors composed of two-dimensional sheets held together by van der Waals interactions, which makes them prime candidates for the intercalation of various guest species.  We review the intercalation chemistry of FeSe and FeS superconductors and discuss their synthesis, structure, and physical properties.  Before we review the latest work in this area, we provide a brief background on the intercalation chemistry of other inorganic materials that exhibit enhanced superconducting properties upon intercalation, which include the transition metal dichalcogenides, fullerenes, and layered cobalt oxides.  From past studies of these intercalated superconductors, we discuss the role of the intercalates in terms of charge doping, structural distortions, and Fermi surface reconstruction.  We also briefly review the physical and chemical properties of the host materials---mackinawite-type FeS and $\beta$-FeSe.
The three types of intercalates for the iron chalcogenides can be placed in three categories:  1.) alkali and alkaline earth cations intercalated through the liquid ammonia technique; 2.) cations intercalated with organic amines such as ethylenediamine; and 3.) layered hydroxides intercalated during hydrothermal conditions. A recurring theme in these studies is the role of the intercalated guest in electron doping the chalcogenide host and in enhancing the two-dimensionality of the electronic structure by spacing the FeSe layers apart.
We end this review discussing possible new avenues in the intercalation chemistry of transition metal monochalcogenides, and the promise of these materials as a unique set of new inorganic two-dimensional systems.  

\vspace{5pt}

\end{abstract}


\end{frontmatter}


\tableofcontents


\section{Introduction}

We review the recent work on the intercalation chemistry of iron chalcogenide superconductors. Given that they exhibit zero resistance for electrical currents below a critical temperature ($T_c$), superconductors hold great promise in our future energy needs.\cite{Carrasco:2006vg, Luongo:2002wf, Ribeiro:2002vl, Larbalestier:2001tx}  Furthermore, superconductors have been proposed for devices that stabilize the electrical power grid by storing energy mechanically in flywheels or electromagnetically in toroidal magnets.\cite{Bray:2008ix, Deng:2008wu}  Much as previously known superconducting materials, the iron-based compounds have a role to play in our future energy needs.\cite{Hosono_2015, Katase:2011dt}

The iron chalcogenide superconductors remain topical due to their versatile solid state chemistry that allows new materials to be discovered and their ability to be isolated as single layers.  Furthermore, the highest $T_c$ so far observed in any iron-based system has been in single layered FeSe where reports vary from 65 K to 100 K.\cite{He_2013, Tan_2013, Ge_2015}  Just like the high-$T_c$ cuprates, the iron-based systems appear to be unconventional in their superconducting mechanism\cite{Paglione_2010, Johnston_2010} and this raises more hope that solid state chemists will continue to make significant discoveries in this field.

\begin{figure}[t!]
\centering
\includegraphics[width=0.90\linewidth]{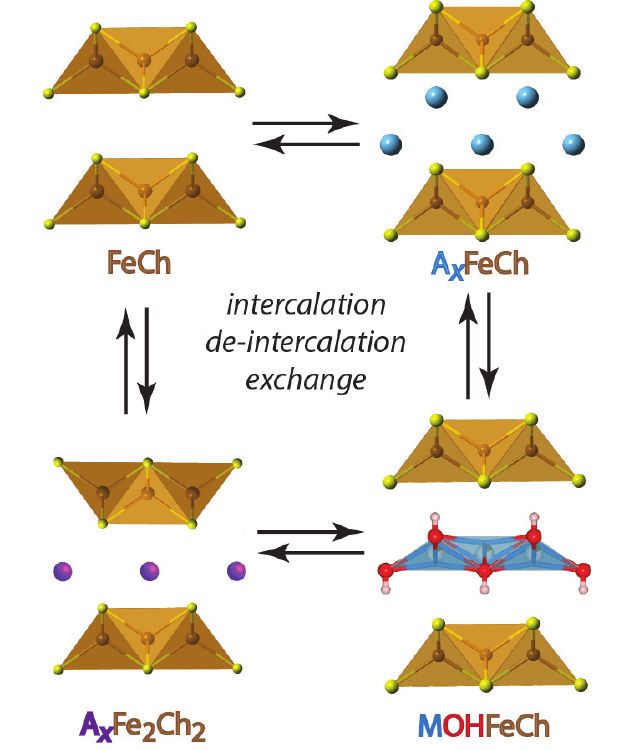}
\caption{The various chemical processes for layered iron chalcogenides (Fe\textit{Ch}) including intercalation of species \textit{A}, where \textit{A} can be a cation, a neutral species such as ammonia, or both.  The various structure types leads to different stacking sequences between the Fe\textit{Ch} layers and therefore different space groups.  The literature on FeSe has revealed that it is possible to generate all of the above structures and that each of the three processes have been observed in the chemistry of iron chalcogenides.}
\label{fig_intro}
\end{figure}

Although similar in many respects to the iron-based pnictides, the chalcogenides do show considerable differences in their physical and chemical properties.  While the arsenide phases are held together by ionic forces between the cationic layers, $e.g.$ (LaO)$^+$ and Ba$^{2+}$, and the anionic (FeAs)$^-$ layers, the chalcogenides can be held by van der Waals interactions alone.  These comparatively weaker interactions make the Fe$Ch$ layers, where $Ch=$S$^{2-}$ and Se$^{2-}$, ideal hosts for intercalation chemistry such as alkali metal insertion along with various other types of guest species such as ammonia.  Since the pnictides do not express intercalation chemistry, we have left them out altogether in this review.  The reader curious about the major differences between chalcogenides and pnictide superconductors can find several reviews written on these two major categories.\cite{Paglione_2010, Johnston_2010, Mizuguchi_2010, Ishida:2009vd, Mou_2011, Ivanovskii_2011} 

We review the chemical techniques for inserting ionic and molecular species into Fe$Ch$ hosts and their resulting physical properties.  An example of the different layered structures we will be reviewing and their intercalation chemistry is presented in the schematic of Figure \ref{fig_intro}.  We will examine three major methods for intercalating guest species, or \textit{intercalates}, which include insertion of 1.) electropositive metals in liquid ammonia, 2.) metals and organic amines, and 3.) extended hydroxides through hydrothermal conditions.  Each method plus post-synthetic treatment can lead to guest-host compounds with different structures as shown in Figure \ref{fig_intro}.  

Before presenting the latest studies on the intercalation of iron-based materials, we briefly review the history of intercalation chemistry of superconductors such as C$_{60}$ and metal dichalcogenides.  It is also instructive to present the crystal chemistry and physical properties of the simple binary iron chalcogenides---FeS, FeSe, and FeTe, which act as the hosts.  The main goals of such chemistry has been to both prepare new compounds and ultimately control the physical properties of the binary compounds in order to discover the underlying mechanism for superconductivity.  We believe the reader will recognize, however, that this chemistry has broader implications than superconductivity and could represent a new area for the preparation of two-dimensional (2D) inorganic materials.

\section{History of intercalation chemistry and superconductivity}
\label{history}

This year, 2016, is a leap year since February has an extra day.  Such an insertion of a day into the calendar is known as an intercalation, and chemistry has borrowed this term to analogously describe the expansion in a solid upon inserting a 'foreign' species.  Intercalation chemistry in inorganic materials has a long history,\cite{Whittingham_1978} and its applications range from electrochemical energy storage\cite{Whittingham_1976} to the sequestration of waste material in layered clays or sulfides.\cite{Manos_2012, Mertz_2013, Lerf_2014}  It is useful to briefly review the terminology in this field and some of the work done on past superconductors.

Like closely related interstitial compounds,\cite{Wells} intercalation compounds are inorganic solids in which guest species such as ions or molecules occupy open crystallographic sites of the host.  Unlike interstitial compounds, however, intercalation hosts usually consist of structural elements such as chains and layers that are held together by weak forces such as those of the van der Waals type.  The guest species, or \textit{intercalate} can cause the unit cell of the host to expand upon insertion but it should not change the crystal structure significantly.  Furthermore, the intercalation and de-intercalation processes should be reversible and possible at temperatures lower than those typical of solid state reactions so that no major rearrangement of the host's structure occurs.  

Given the constraints of the intercalation processes, two dimensional (2D) materials are the most common hosts.\cite{Lerf_2014}  Materials with one-dimensional (1D) motifs can facilitate exit and entry of guest species but also undergo major re-organization upon intercalation.  Conversely, three-dimensional (3D) materials retain their structural integrity but impede facile intercalation of guests.  Classic inorganic solids in the 2D category include layered materials such as graphite,\cite{Dresselhaus_2002} graphite oxide,\cite{Buchsteiner_2006} and montmorillonite -- a silicate clay.\cite{Usuki_1993}  If the host is redox active, then the intercalation process is either reductive or oxidative, and this special property has been explored in electrochemical devices such as rechargeable batteries.\cite{Whittingham_1976, Goodenough_2013}

In order to raise some key questions from the literature on iron chalcogenides, we briefly discuss three other inorganic hosts that enhance their superconducting properties upon intercalation.  These include the layered metal dichalcogenides, layered cobaltates, and three-dimensional fullerene crystals.  These cases present to us the importance of charge doping, Lewis acid-base chemistry, and structure-property relationships on superconductivity, and guide our understanding of iron chalcogenide intercalation chemistry.

\subsection{Transition metal dichalcogenides}
\label{MCh2}

\begin{figure}[!t]
\centering
\includegraphics[width=0.80\linewidth]{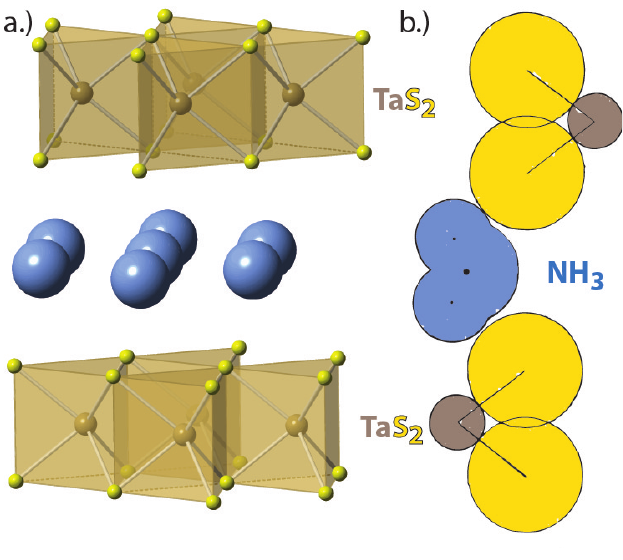}
\caption{a.) The crystal structure of TaS$_2$$\cdot$(NH$_3$) where the nitrogen atoms (blue) are in a hexagonal arrangement between the sulfide sheets. Crystal structure from Ref. \cite{Chianelli_1975} b.) A representation of the orientation of the ammonia molecule so that the 3-fold axis is oriented parallel to the TaS$_2$ sheets.  Reprinted with permission from Ref. \cite{Gamble_1975}. Copyright 2008, AIP Publishing LLC.}
\label{fig_NH3-TaS2}
\end{figure}

Metal dichalcogenides are redox active hosts with a van der Waals gap between the layers that allows facile intercalation.\cite{Friend_1987}  Furthermore, they are semiconducting, which aids in the insertion and removal of cations by conducting charge-compensating carriers (\textit{i.e.} electrons and/or holes).  In the early 1970's, several groups demonstrated that upon intercalation, the superconducting critical temperature, or $T_c$, could be raised in dichalcogenides such as TaS$_2$ and NbSe$_2$.\cite{Gamble_1970, Salvo_1971, Geballe_1971, Murphy_1975}  In particular, the work of Geballe et al. demonstrated how a large array of organic and inorganic hosts could act as Lewis bases, or electron donors, and therefore easily intercalate into the layered materials that act as Lewis acids.\cite{Gamble_1970}  The layered hosts can accept Lewis bases such as ammonia, pyridine, or even amides with long aliphatic chains such as stearamide.  Since the metal dichalcogenides act as Lewis acids, the $T_c$'s of the final adducts were related to relevant parameters such as interlayer spacing and the $pK_A$'s of the Lewis bases.  

Lerf has raised doubts, however, whether Lewis acid-base chemistry truly drives the intercalation process in dichalcogenides.\cite{Lerf_2014}  The chief objection from Lerf concerns the findings from crystallographic studies showing that the orientation of the NH$_3$ molecule within the layers is such that its 3-fold axis is oriented along the layers as opposed to perpendicular to them as shown in Figure \ref{fig_NH3-TaS2} for the case of TaS$_2$.\cite{Gamble_1975}  Similarly, the 2-fold axis of the pyridine ring is oriented parallel to the dichalcogenide layers.\cite{Riekel_1979}  This would imply that the lone pair on the nitrogen atom is not oriented towards the layers and is therefore ineffective as a Lewis base. 

Besides the electron donating properties upon intercalation, structure-property relationships are also important.  The common structural motif is the 2D layer formed by infinite edge-sharing $MCh_6$ polyhedra where $M$ = transition metal and $Ch$ = S, Se, and Te.  Polytypism, is common in these materials, since the coordination of the transition metal can be either octahedral or trigonal prismatic due to changes in the ion stacking sequence along the interlayer direction (Figure \ref{fig_NH3-TaS2}).  However, only when the metal is in trigonal prismatic coordination has superconductivity been observed.\cite{Friend_1987}  In their studies of 50 different intercalates of various sizes for TaS$_2$, which increased the interlayer spacing anywhere between 3 \AA \, to 52 \AA\,, Gamble et al. found the $T_c$ bears little correlation to interlayer spacing.\cite{Gamble_1970}  Instead, they attributed any enhancement in $T_c$ to the efficacy of charge transfer from the intercalate to the host.

Alkali metals can also be intercalated into dichalcogenide hosts, either in a liquid ammonia solution or electrochemically, which allows for a more direct measure of electron doping than the Lewis basicity of amines.\cite{Friend_1987}  Furthermore, Sch\"ollhorn,\cite{Schollhorn_1975, Lerf_1977, Schollhorn_1977, Schollhorn_1979} Colombet,\cite{Colombet_1989} and others\cite{McMillan_1991} have argued that the intercalation of amines actually occurs through the oxidation of such species so that the true nature of the intercalates is  the amine plus its conjugate acid (\textit{e.g.} NH$_3$ and NH$_4^+$), which would also charge dope the dichalcogenide host.  The formation of cationic intercalates, in fact, has been suggested as the driving force for the intercalation of such species into the host.\cite{Colombet_1989}  Once the cationic form of the amines has been inserted, the other amines orient themselves in such a way to direct the lone pair of electrons on the nitrogen atoms towards the cations.

\subsection{Other intercalated superconductors}
\label{cobaltates}

\begin{figure}[!t]
\centering
\includegraphics[width=1.00\linewidth]{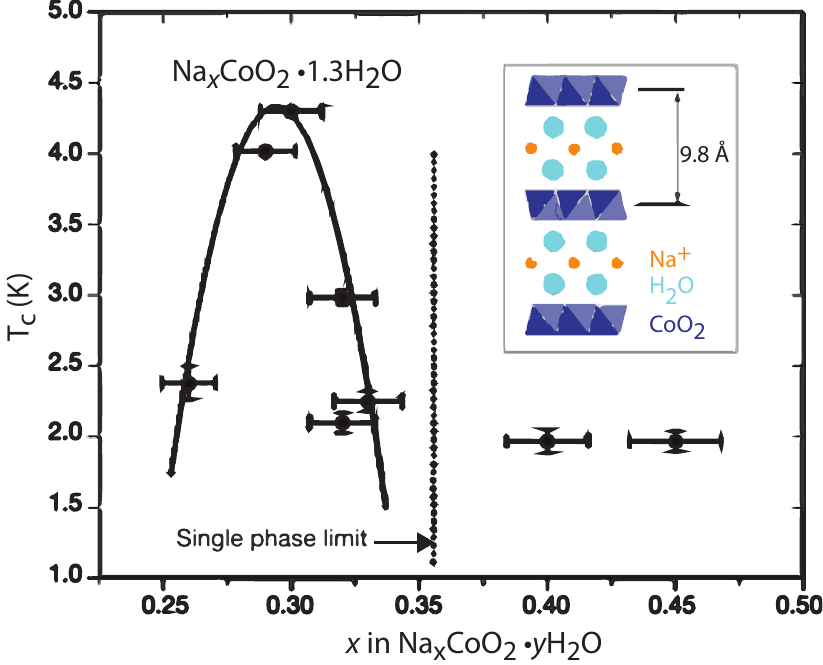}
\caption{The phase diagram for the layered oxides, Na$_x$CoO$_2$$\cdot$ 1.3 H$_2$O as a function of sodium intercalation.  The maximum $T_c$ is reached for nearly one-third doping and for the hydrated version.  Figure is adapted from Ref. \cite{Schaak:2003aa} by permission from Macmillan Publishers Ltd: Nature, copyright 2003.}
\label{fig_Na-CoO2}
\end{figure}

A more recent example of intercalation chemistry than that of the dichalcogenides includes the layered Na$_x$CoO$_2$ materials.\cite{Takada:2003aa}  These cobalt oxides consists of CoO$_6$ octahedra that edge-share to form layers of the brucite-type structure.  Na and Li cations can fill in the interstitial positions in between and the shuttling of these cations has made the cobaltates an important cathode material for Li-rechargeable batteries.\cite{Goodenough_2013}  In the sodium intercalated case, when the right amount of Na$^+$ is present along with water molecules, superconductivity is induced.  Interestingly, the anhydrous case is non-superconducting, and the highest $T_c$ of 4.5 K is found for the stoichiometry of Na$_{0.3}$CoO$_2$ $\cdot$ 1.3 H$_2$O  (see Figure \ref{fig_Na-CoO2}).\cite{Schaak:2003aa}   At values for $x$ greater than $\frac{1}{3}$, anhydrous Na$_x$CoO$_2$ can exhibit various ordered states for its charge and spin degrees of freedom.\cite{Huang_2004}  These structures, much like the metal dichalcogenides, demonstrate that they can can accommodate more than one type of guest simultaneously, and that this can be advantageous for superconductivity.

Another historical example is that of the intercalated fullerenes.  Just as graphite can be easily intercalated,\cite{Dresselhaus_2002} the molecular allotrope of carbon, C$_{60}$ can also accommodate metal cations within its interstices.  In the 1990's, researchers found that when fullerenes are charge-doped to become fullerides, they are tuned from semiconducting materials to metals and superconductors.\cite{Stephens:1991aa, Diederich_1991}  The molecular nature of the crystal and the relatively weak interaction between the C$_{60}$ `hard spheres' make intercalation possible not only for alkali metals but also molecular species such as ammonia.\cite{Rosseinsky:1993wc}  A notable difference in the intercalation chemistry of C$_{60}$ and graphite is the former's preference for reductive intercalation whereas graphite can undergo both oxidative and reductive insertion.\cite{Rosseinsky:1995th, Rosseinsky:1998vb} Remarkably, the intercalated fullerenes such as  K$_3$C$_{60}$, Rb$_3$C$_{60}$, and Rb$_2$CsC$_{60}$ exhibit relatively high $T_c$'s, up to 32 K.\cite{Rosseinsky:1998vb}. 

The historical examples given above illustrate some of the key factors to consider for the iron-based superconductors.  These considerations include the role of the intercalate in charge doping into the density of states near the Fermi level, distorting the relevant structural parameters, and changing the dimensional nature of the Fermi surface.   Nevertheless, all of these examples are classified as conventional superconductors whose underlying physics is satisfactorily described by Bardeen-Cooper-Schrieffer (BCS) theory.\cite{Friend_1987, Murphy_1975, Harper_1977, Rosseinsky:1998vb} Therefore not all lessons learned from these studies will apply to the iron-based materials.


\section{Chemistry and physics of the Fe\textit{Ch} hosts}

Before discussing the intercalation chemistry of the iron based superconductors, it is instructive to describe the structure, bonding, and electronic properties of the host materials.  The Fe$Ch$ ($Ch$=S,Se,Te) layers, like the transition metal dichalcogenides, are held by weak van der Waals forces that make them susceptible to intercalation.  In the case of the iron telluride, however, the van der Waals gap is too small on account of the large anionic radius of Te$^{2-}$ (2.21 \AA),\cite{Shannon:1976vx} and this hinders any significant intercalation chemistry.  Therefore, this review is focused on the selenides and sulfides, but we do present the electronic and physical properties of Fe$_{1+x}$Te (where $x$ represents excess iron) briefly in order to compare and contrast with those of FeS and FeSe.

\begin{figure}[t]
\centering
\includegraphics[width=0.90\linewidth]{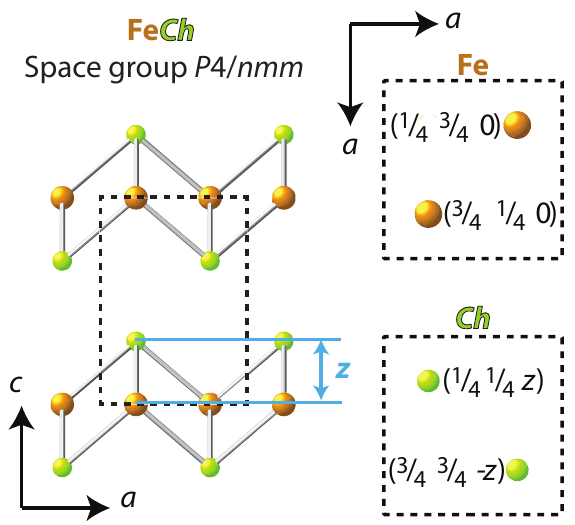}
\caption{The crystal structure of the simple binary Fe$Ch$ with the anti-PbO type structure.  Setting 2 for the origin is represented for this tetragonal space group.  The dashed line represents the unit cell, and $z$ is the so-called anion height.}
\label{fig_FeCh}
\end{figure}

\subsection{Crystal structures}

Although many binary iron sulfides and selenide phases are known, we are only interested in those that crystallize in the anti-PbO type structure. As shown in Figure \ref{fig_FeCh}, the structure is described by the tetragonal space group $P4/nmm$ (no. 129), which has two settings.  In setting 1, the iron atoms are located at the origin of the cell while in setting 2 the origin is located at the center of inversion symmetry with the iron atoms displaced ($\frac{3}{4}$,$\frac{1}{4}$, 0) from the origin.\cite{Wilson_2010}   Figure \ref{fig_FeCh} shows setting 2.  Either way, the site symmetries of the atoms are such that the Fe cations have $\overline{4}m2$ site symmetry and the $Ch^{2-}$ anions $4mm$ symmetry. Furthermore, the $n$-glide normal to the $c$-axis leads to the adjacent layers being staggered by ($\frac{1}{2}, \frac{1}{2}$) in the $ab$-plane, which maximizes the distance between the $Ch^{2-}$ anions.  In the resulting two-dimensional square lattice of Fe cations, each square is capped by a chalcogenide anion, either above or below, which leads to tetrahedral coordination for each iron cation. 

\begin{figure}[!t]
\centering
\includegraphics[width=1.00\linewidth]{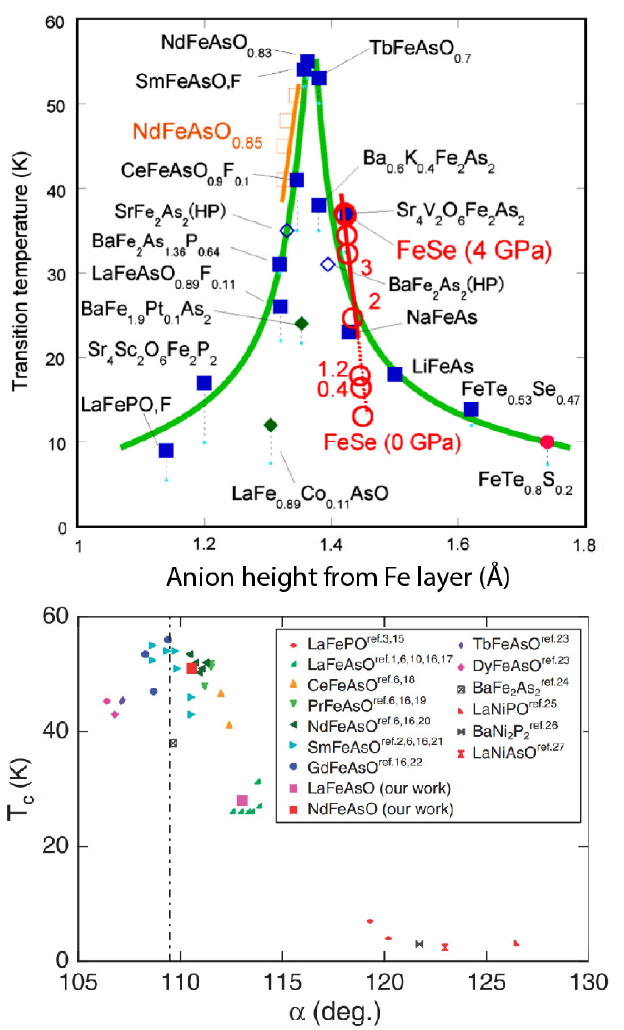}
\caption{Top: The correlation of $T_c$ with the anion height parameter for various members of the iron arsenide and chalcogenide superconductors.  Reprinted with permission from Ref. \cite{Mizuguchi:2010vm}. Copyrighted by the Physical Society of Japan. Bottom: The correlation of $T_c$ with the tetrahedral bond angle $\alpha$ for the iron arsenide family of superconductors.  Reprinted with permission from Ref. \cite{Lee_2008}. Copyright IOP Publishing.  Reproduced with permission.  All rights reserved.}
\label{fig_StrucTrends}
\end{figure}

The only crystallographic parameters that can be manipulated include the lattice constants and the $Ch$ anion $z$-parameter, also know as the anion height from the iron square lattice.  These two parameters in addition to the lattice constants are therefore enough to describe important properties such as Fe--Se bond distances, Fe--Fe distances, which influence the $d$-orbital dispersion bands, and the $Ch$--Fe--$Ch$ bond angle.  As shown in Figure \ref{fig_StrucTrends}, the anion height and the variance of the tetrahedral bond angle both correlate with $T_c$ across the various iron-based superconductor families.\cite{Lee_2008, Mizuguchi:2010vm, Wilson_2010, Lee_2012, Horigane:2009ky}   

Structurally, the layered iron chalcogenides provide a nice contrast to the transition metal dichalcogenides.  The former is composed of a square planar lattice of $d$-block metals whereas the latter is composed of a hexagonal array.  In the iron-based chalcogenides, 2D layers consist of edge-sharing tetrahedra (Figure \ref{fig_FeCh}) whereas in the dichalcogenides $MCh_6$ trigonal prisms or octahedra complete the layers.  These two structural distinctions lead to differences in both the local electronic properties determined by crystal field splitting considerations and to the long-range electronic properties determined by the band structures. 

\begin{figure}[t]
\centering
\includegraphics[width=1.00\linewidth]{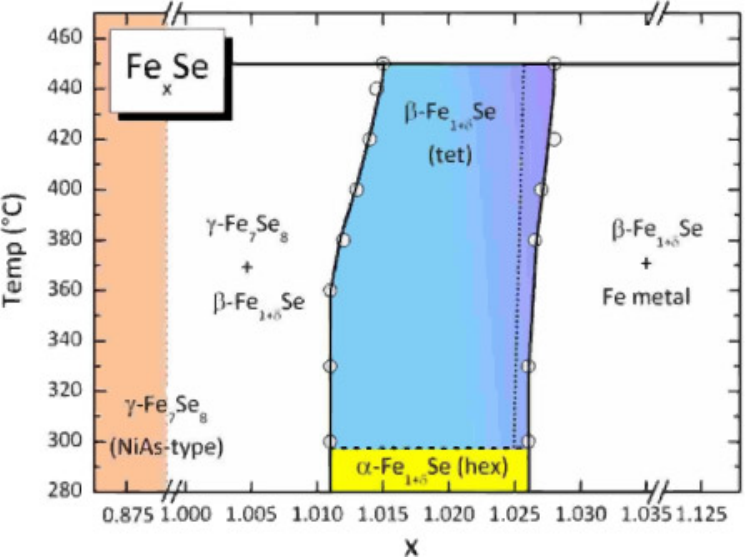}
\caption{The binary phase diagram of iron and selenium, with the phase space for $\beta$-FeSe shown in blue.  Reprinted figure with permission from Ref. \cite{McQueen_2009}. Copyright 2009 by the American Physical Society.}
\label{fig_PhaseDia-FeSe}
\end{figure}

One of the most studied iron-based superconductors is iron selenide.  In its anti-PbO structure it is known as $\beta$-FeSe.  This phase exists in a narrow window of stability in the iron-selenium binary phase diagram (300 K$ < T < $ 450 K),\cite{McQueen_2009} as shown in Figure \ref{fig_PhaseDia-FeSe}, and its preparation has mostly consisted of powders through direct reaction of the constituent elements.  Polycrystalline FeSe, however, can be prepared through hydrothermal routes, and care must be taken to prevent oxidation, which is detrimental to the superconducting properties.\cite{Greenfield_2015b}  Since this phase does not melt congruently, single crystal growth has been challenging.   Several groups have found chemical vapor transport and salt flux methods, however, as successful techniques for the preparation of high quality single crystals.\cite{Hara2010} 

When near stoichiometric, FeSe superconducts below 8 K.\cite{Hsu_2008, Kotegawa_2008} Interestingly, under high pressure the $T_c$ increases up to 37 K.\cite{Margad_2008, Imai_2009}   Even higher $T_c$'s have been observed in FeSe in isolated single layers.  Through \textit{in situ} electrical resistivity measurements during photoelectron spectroscopy measurements, $T_c$'s have ranged from 65 K to 100 K in single layered FeSe.\cite{He_2013, Tan_2013, Ge_2015}  This is the record for any iron-based superconductor, and it raises the possibility that materials chemists and physicists could develop a new family of high-$T_c$ superconductors. 

Above $T_c$ FeSe undergoes a crystallographic phase transition that may provide clues to its superconducting pairing mechanism.  Through various diffraction studies,\cite{Hsu_2008, McQueen_2009b, Greenfield_2015b} a transition from $P4/nmm$ to orthorhombic $Cmmm$ at 90 K in superconducting FeSe was observed.  This lowering of symmetry is accomplished through the breaking of the four-fold axis and the $n$-glide plane normal to the $c$-direction. However, for the non-superconducting Fe$_{1.03}$Se, which contains extra interstitial iron, the symmetry lowering is not observed, and the driving mechanism lifting the degeneracy of the $xy$-plane was proposed to not be magnetically driven.\cite{McQueen_2009b}  More recent diffraction studies by Pachmayr et al.\cite{Johrendt_2016} have found that FeSe prepared by hydrothermal conditions as opposed to traditional solid state techniques actually leads to a triclinic system and a lower temperature of transition near 60 K.  Systematic diffraction studies including neutrons at lower temperatures of various FeSe single crystal samples would help shed further light on how composition drives the symmetry lowering transition.  Shear-modulus\cite{Bomer_2015} and angle resolved photoemission spectroscopy (ARPES) studies have proposed that the orthorhombic distortion leads to a so-called nematic phase where the stripes of elongation are driven by an orbital-ordering of the $d$-states.\cite{Watson_2015}  However, this topic remains unsettled with new results from neutron studies\cite{Rahn_2015} and theoretical modeling\cite{Glasbrenner_2015, Wang_2015} that report the possible importance of spin fluctuations in FeSe. 

\subsection{$\beta$-FeSe}

\subsection{Mackinawite FeS}

In contrast to $\beta$-FeSe, the sulfide analogue does not exist in the thermodynamic phase diagram of iron sulfides.  Within the binary phase diagram of the iron sulfur system, there are approximately 7 different compositions with 10 different crystallographic structures.\cite{Waldner_2005, Vaughan_1971} Many of the phases are well known minerals such as pyrite (FeS$_2$) or pyrrhotite (Fe$_{1-x}$S).  Therefore tetragonal FeS with the anti-PbO type structure, also known as mackinawite, is a metastable phase.

In the early crystallographic studies, the diffraction data set was not of high quality enough due to sample broadening, and the $z$-parameter of the S$^{2-}$ anion was fixed to $z=1/4$.\cite{Berner_1964}    Subsequent studies by Lennie et al. found $z$ to be larger than 0.25 in order for the FeS$_4$ tetrahedron to have more regular tetrahedral angles.\cite{Lennie:2009ta}  The various thermal studies by Lennie et al. found that while well crystallized mackinawite eventually decomposes to pyrotite near 530 - 545 K,\cite{Lennie:1995wy} powder samples can first convert to greigite Fe$_3$S$_4$ at a much lower temperature of 373 K.\cite{Lennie_1997}  Therefore, it is likely that the larger particle sizes and therefore samples with smaller surface areas have a higher barrier towards conversion of tetragonal FeS to one of its thermodynamically stable phases.

FeS is typically prepared by solution routes either under ambient conditions\cite{Berner_1964, Lennie:2009ta, Rickard_2006} or hydrothermally.\cite{Sines_2012, Lai_2015, Borg_2016}  As a polycrystalline material, mackinawite can be prepared by oxidation of iron metal in an acetic acid/acetate buffer solution.  Afterwards, an aqueous solution of Na$_2$S hydrate is added to the buffer leading to the rapid precipitation of tetragonal FeS.\cite{Lennie:2009ta}  According to the Pourbaix diagram of iron, under basic conditions, Fe$^{3+}$ species predominate whereas Fe$^{2+}$ predominate at lower pH's.\cite{Marie_1998, Drissi_1999}  Therefore, in the initial low-pH solution, the iron is oxidized to evolve Fe$^{2+}$ species, and the buffer serves to minimize the amount of Fe$^{3+}$ species upon adding the sodium sulfide.  Lennie et al. have argued that mackinawite is kinetically stabilized and transforms to other phases only through oxidation or further sulfurization.\cite{Lennie:1995wy}

Studies on the physical properties of mackinawite FeS have been inconsistent with some studies reporting a semiconducting ground state\cite{Sines_2012, Denholme_2014, Denholme_2014b} and others superconductivity.\cite{Lai_2015, Borg_2016}  Nanocrystalline FeS, prepared by the crystallization of an amorphous black solid through a solvothermal method, was found to be semiconducting and ferrimagnetic.\cite{Sines_2012}  Likewise Denholme et al. found their mackinawite to be semiconducting\cite{Denholme_2014} but upon the application of pressure, the materials become more metallic.\cite{Denholme_2014b}  Denholme et al. suggest that mackinawite may indeed be metallic, and that the semiconducting behavior arises from grain boundaries in the polycrystalline samples.  Ferrimagnetism was also observed in the magnetization vs. applied field curves, and the magnetic susceptibility vs. temperature curves indicate that the N\`eel temperature is above room temperature.

A much earlier study, however, had found no long range magnetic ordering in mackinawite.  Bertaut performed neutron diffraction studies on powder samples of mackinawite.\cite{Bertaut_1965}  In the powder pattern at 10 K, antiferromagnetic Bragg peaks were not observed. Bertaut concluded that FeS is not antiferromagnetic and cannot therefore be ferrimagnetic.  Furthermore, electronic structure calculations had predicted that, much like its heavier analogues FeSe and FeTe, mackinawite should be a metal or semimetal.\cite{Kwon_2011, Subedi_2008, Brgoch_2012}  The observation by Lennie et al. that the surface of mackinawite first converts to Fe$_3$S$_4$\cite{Lennie_1997} could explain the ferrimagnetic behavior in magnetization measurements since Fe$_3$S$_4$, which has the spinel-type structure, is a ferrimagnetic semiconductor.\cite{Roberts:1995aa}

\begin{figure}[!t]
\centering
\includegraphics[width=0.90\linewidth]{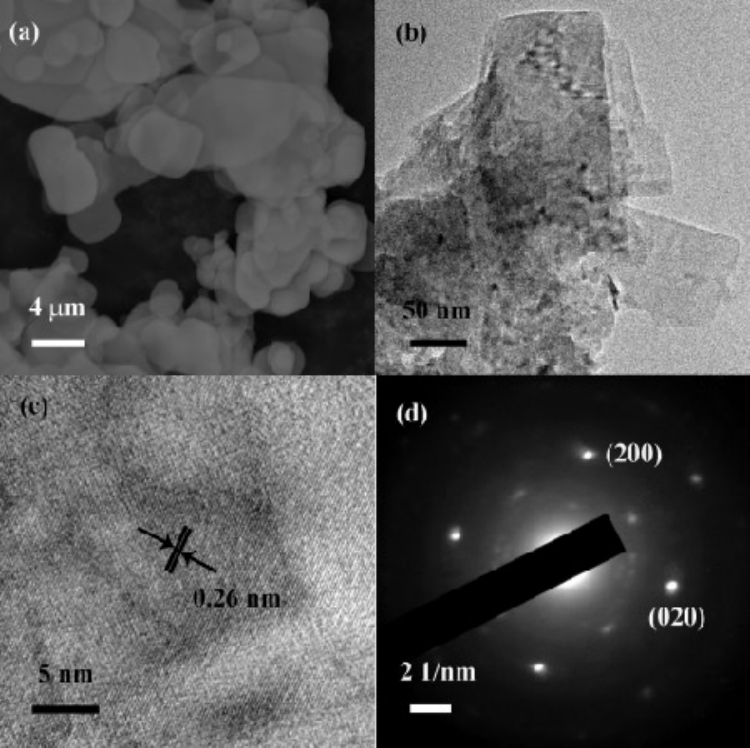}
\caption{Electron microscopy images of superconducting FeS including a.) SEM, b.) TEM, c.) high-resolution TEM, and d.) selected area electron diffraction.  Reprinted with permission from Ref. \cite{Lai_2015}. Copyright 2015 American Chemical Society.}
\label{fig_TEM-FeS}
\end{figure}

Very recently, Lai et al. found mackinawite FeS to be metallic down to 5 K, below which it becomes superconducting.\cite{Lai_2015}  Lai et al. prepared a high quality sample of tetragonal FeS by oxidizing iron metal under hydrothermal conditions in the presence of Na$_2$S hydrate (Figure \ref{fig_TEM-FeS}).  Likewise, Borg et al. demonstrated that single crystals of K$_x$Fe$_{2-y}$S$_2$ could be de-intercalated hydrothermally to prepare single crystals of FeS.\cite{Borg_2016} The key to obtaining superconducting crystals was to maintain the solution under basic conditions and to provide enough iron powder during the hydrothermal synthesis in order to fill in the vacancies in the K$_x$Fe$_{2-y}$S$_2$ samples. The single crystals afforded the true ground state properties of mackinawite, which include highly anisotropic superconducting and normal state properties.  This high anisotropy is indicative of the 2D electronic structure of FeS.  Above $T_c$, FeS is truly a metal with weak Paul paramagnetism.\cite{Borg_2016}


\subsection{Fe$_{1+x}$Te}
 
While Fe$_{1+x}$Te is an interesting 'parent' phase (\textit{i.e. not superconducting until doped}),\cite{Fang:2008uz, Mizuguchi:2009va, Bao:2009vl, Zajdel:2010vt, Liu:2010dc, Bhatia:2011gd} we have not seen any literature on its intercalation chemistry.  The difficulty may arise for several reasons including the smaller van der Waals gap due to the larger Te$^{2-}$ anionic radius with respect to its lighter congeners.  Furthermore, intercalation often occurs under highly basic conditions, and Te is more easily oxidized under these conditions than Se or S since it is more electropositive.  De-intercalation chemistry on Fe$_{1+x}$Te and  Fe$_{1+x}$Te$_{0.7}$Se$_{0.3}$, however, has been demonstrated in order to remove the interstitial iron, $x$,\cite{Rodriguez:2010aa, Rodriguez:2011kb, Stock:2011jp} which has large implications for the magnetic, transport, and superconducting properties.\cite{Rodriguez:2011ew, Rodriguez:2013cb, Stock:2014aa}.  Therefore, the iron telluride system may still be an interesting host structure to explore for possible intercalation, ion exchange, and de-intercalation chemistry.


\subsection{Electronic structure of the hosts}

\begin{figure}[!t]
\centering
\includegraphics[width=0.70\linewidth]{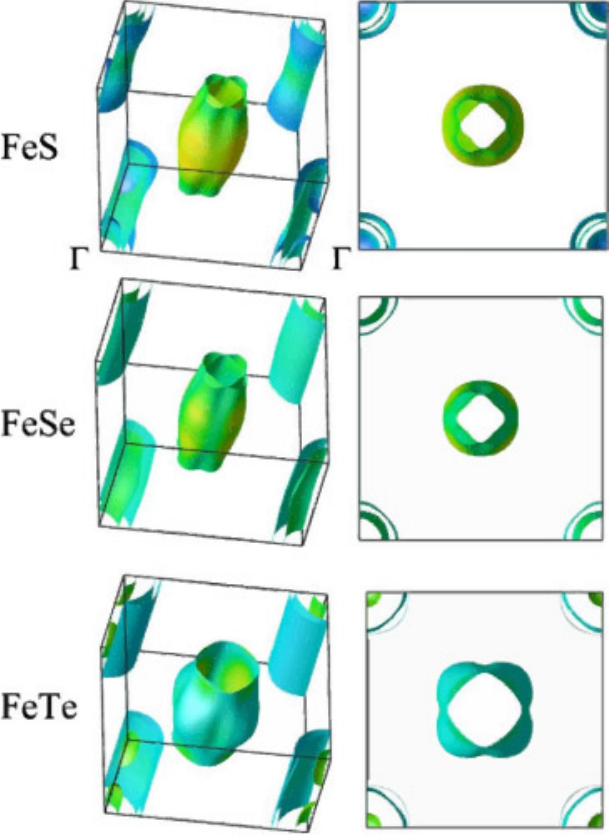}
\caption{The Fermi surfaces the binary iron chalcogenides with the anti-PbO type structure.  The zone-centered and corner-centered surfaces are hole and electron pockets, respectively.  The cylindrical aspect of these surfaces demonstrate the two-dimensional aspect of their electronic structure.  Reprinted figure with permission from Ref. \cite{Subedi_2008}. Copyright 2008 by the American Physical Society.}
\label{fig_FeChFermiSurf}
\end{figure}

Finally, it is important to discuss the electronic structure of the host structures.  The square Fe sublattice has a strong 2D character in its electronic structure.  The Fe$Ch$ layers are metallic in nature, and the $d$-states of the Fe cations form bands near the Fermi level.  Unlike the previous intercalated superconductors, the iron-based materials are multi-band superconductors.  The 2D character is clearly observed in the cylindrical nature of the Fermi surfaces as found by density functional theory (DFT) calculations (Figure \ref{fig_FeChFermiSurf}).\cite{Subedi_2008}  In this respect, the chalcogenides do have some similarities to the iron pnictides, where nesting between the electron and hole pockets leads to a spin density wave instability and crystallographic phase transitions.\cite{Johnston_2010}  Upon charge doping, the SDW vanishes and the pnictides become superconducting.\cite{Paglione_2010}  While SDW order has been found in the Fe$_{1+x}$Te phases, no such order has been observed in FeS and FeSe. 

The Fermi surface of FeS and FeSe show several separate sets of bands: one set is zone-centered and are hole carrying, and the other set is corner-centered and electron carrying.\cite{Subedi_2008}  The dispersion curves reveal that these bands consist of the various Fe $d$-orbitals while the $Ch$ $p$-orbitals are buried deep in energy.  According to a recent ARPES study of FeSe, the bands around the zone center consist largely of the $d_{xz}$ and $d_{yz}$ orbitals, which are degenerate in the tetragonal phase, but split upon the orthorhombic distortion near 90 K.\cite{Watson_2015}  The electron pocket meanwhile is dominated by the $d_{xy}$ orbital.  From their ARPES study on high quality crystals, Watson et al. conclude that orbital-charge ordering of the $d_{xz}$ and $d_{yz}$ bands drive the crystallographic phase transition near 90 K in FeSe.\cite{Watson_2015} 

A simplistic way to understand the band structure from a chemical bonding perspective is to consider the ligand field splitting for the FeSe$_4$ tetrahedra.  For tetrahedral coordination, the doubly degenerate $e$ states consist of the $d_{x^2-y^2}$ and $d_{z^2}$ orbitals.  The $e$-states are lower in energy than the triply-degenerate $t_2$-states, which consist of the $d_{xz}$, $d_{yz}$, and $d_{xy}$ orbitals.  In the case of the layered Fe$Ch$ materials, rather than being antibonding the $t_2$-states are mostly nonbonding.  The local geometry around the iron cations, however is not ideally tetrahedral since the site symmetry of the cations is $\overline{4}m2$ (or $D_{2d}$) and therefore lower than $\overline{4}3m$ (or $T_d$).  This site symmetry lifts the degeneracy of the $t_2$-manifold, so that the $d_{xz}$, $d_{yz}$ orbitals form their own degenerate pair.  Perhaps not coincidentally, a reliable structural descriptor for the $T_c$ in iron-based superconductors is the closeness of the tetrahedral bond angles to the ideal 109.5$^{\circ}$, with the highest observed for the arsenide family being $T_c$ (Figure \ref{fig_StrucTrends}b).


\section{Metal intercalation with liquid ammonia}

\begin{figure}[t!]
\centering
\includegraphics[width=0.90\linewidth]{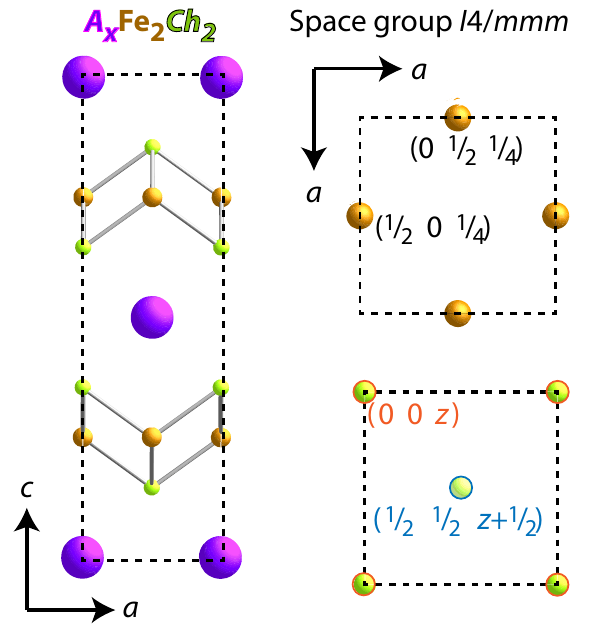}
\caption{The crystal structure of $A_x$Fe$_{2-y}$Se$_2$ where $A$ can be various species including alkali metal cations or moieties such as NH$_3$ and NH$_2^-$.  The structure is also known as the ThCr$_2$Si$_2$-type structure.  }
\label{fig_AFe2Ch2}
\end{figure}

Before discussing the first attempts at intercalating FeSe and FeS, we briefly mention that compounds of the type $A_x$Fe$_{2-y}$Se$_2$ had been prepared before through solid state techniques (where $A^+$ = alkali metal or Tl$^{1+}$ cations).\cite{Guo_2010, Fang_2011, Hu2011, Lei:2011jz, Lei:2011kt, Li_2011, Wang_2011}  These materials exhibit a significant increase in the $T_c$ up to approximately 30 K.  Since single crystals can readily be grown from congruent melts,\cite{Wang2011, Ying_2011} we do not consider their preparation as intercalation chemistry.  We briefly describe them, however, since their crystal structures are related to the intercalated iron chalcogenides and because their shortcomings highlight the need for intercalation chemistry to properly study the superconducting properties of FeSe.

In contrast to the host structure, the alkali metal intercalated phases crystallize into the ThCr$_2$Si$_2$-type structure with a body-centered tetragonal cell and space group $I4/mmm$ as shown in Figure \ref{fig_AFe2Ch2}.  Therefore the relationship between the two adjacent Fe$Ch$ layers is different from that of the host.  In the latter, the layers are staggered with respect to one another by the $n$-glide plane ($\frac{1}{2}+ x,  \frac{1}{2}+y, \overline{z}$).  This is not the case in the body centered cell, which is missing the $n$-glide, and the $Ch^{2-}$ anions are located directly over one another.  Presumably, ionic forces now dominate the bonding between the layers as intercalation of the $A^+$ cations reduces the Fe$Ch$ to become anionic (Fe$Ch$)$^{\delta-}$ layers.

A serious drawback for these ternary selenides, however, is that the $A_x$Fe$_{2-y}$Se$_2$ superconducting phase coexist with an insulating phase,\cite{Hu2011, Wang_2011b, Li_2011b, Liu2011}  which was later found to be near the stoichiometry of $A_2$Fe$_4$Se$_5$.\cite{Pomjakushin_2011, Shoemaker_2012, Ding_2013, Zavalij_2011} Since the superconducting and antiferromagnetic phases are structurally related, they often coexist when prepared by congruently melting the substituent elements.  Shoemaker et al. demonstrated through their careful diffraction studies that the minority phase in the samples is superconducting,\cite{Shoemaker_2012} and several studies have shown that the amount of $x$ leading to superconductivity ranges from 0.3 to 1.0.\cite{Wang_2014, Ying_2011, Guo_2010, Zhang:2011aa}  Intercalation at low temperatures into pristine FeSe layers therefore offer an opportunity to enhance superconductivity and avoid the iron vacancy-populated phase.

While several groups have proposed that the stoichiometry of the superconducting phase to be close to $A_x$Fe$_2$Se$_2$,\cite{Shoemaker_2012} Ding et al.  formulated the superconducting phase from their careful electron microscopy studies as approximately K$_2$Fe$_7$Se$_8$.\cite{Ding_2013}  Therefore, it is the particular ordering of vacancies in $A_2$Fe$_4$Se$_5$ that is thought to be detrimental to superconductivity, favoring instead long-range antiferromagnetic ordering at 559 K.\cite{Bao_2011}  Svitylk et al. further demonstrated that vacancy ordering in itself may not be detrimental by finding superconductivity in the Rb$_x$Fe$_{1-y}$Se$_2$ system, which showed some iron vacancies.\cite{Dmitriev_2011} Interestingly, the magnetic moment per iron cation of 3.31 $\mu_B$ in K$_2$Fe$_4$Se$_5$ suggests that the $d^6$ cations in the tetrahedral crystal field may have an orbital contribution in addition to spin.\cite{Bao_2011}
 
\begin{figure}[t!]
\centering
\includegraphics[width=0.70\linewidth]{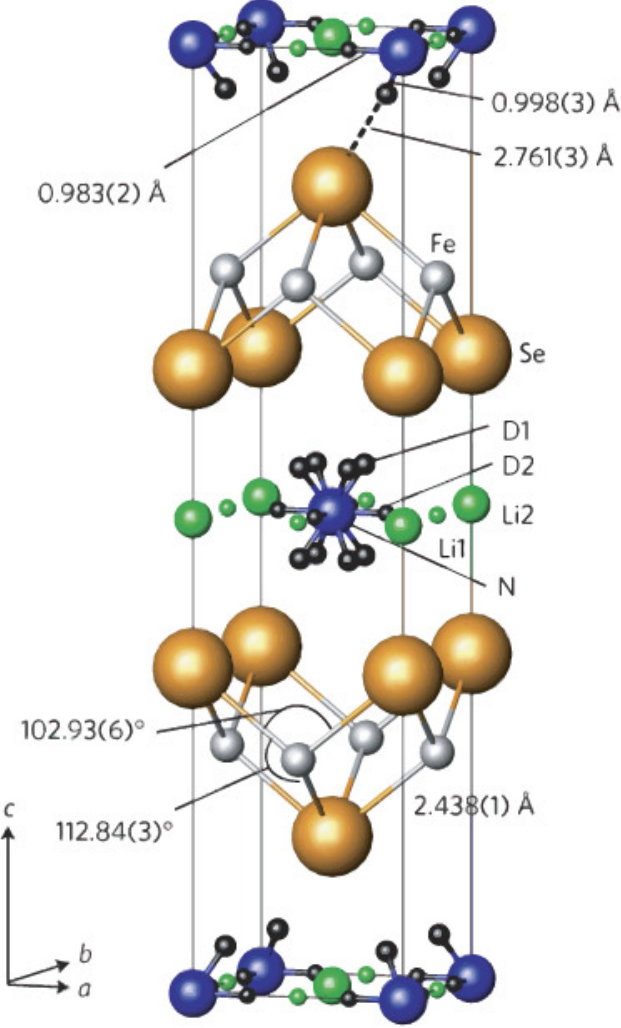}
\caption{The structure of Li$_{0.6}$(ND$_2$)$_{0.2}$(ND$_3$)$_{0.8}$Fe$_2$Se$_2$ where the Li cations reside mostly between the Se$^{2-}$ anions and the ND$_3$ and ND$_2^-$ moieties at the cell center. Reprinted from Ref. \cite{Burrard_2013} by permission from Macmillan Publishers Ltd: Nature, copyright 2012.}
\label{fig_LiND3-Fe2Se2}
\end{figure}

In a similar fashion to the intercalation of alkali metals into dichalcogenides, FeSe was reduced by an alkali metal-liquid ammonia solution, which contains solvated electrons (hence its blue color).\cite{Scheidt_2012, Tian_2013}  Through this ammonia solution alkali and alkaline earth metals can be inserted into the FeSe host and the $T_c$ raised to 43(1) K.\cite{Burrard_2013}  Burrard-Lucas et al. have definitively demonstrated through neutron diffraction studies that this technique also intercalates ammonia and amide (NH$_2^-$) molecules along with Li$^+$ cations in the interstitial positions to form Li$_{0.6}$(NH$_2$)$_y$(NH$_3$)$_{1-y}$Fe$_2$Se$_2$.  Unlike the typical ThCr$_2$Si$_2$-type structure, however, the Li cations are predominately in the $2b$ Wyckoff position, which is not at the unit cell's body center but between the Se$^{2-}$ anions as shown in Figure \ref{fig_LiND3-Fe2Se2}.  Using deuterated ammonia for the neutron experiments, Burrard-Lucas found the ND$_3$ and ND$_2^-$ moieties to be located at the cell center, and the deuterium atoms disordered at the $16m$ site.\cite{Burrard_2013}

\begin{figure}[t!]
\centering
\includegraphics[width=1.00\linewidth]{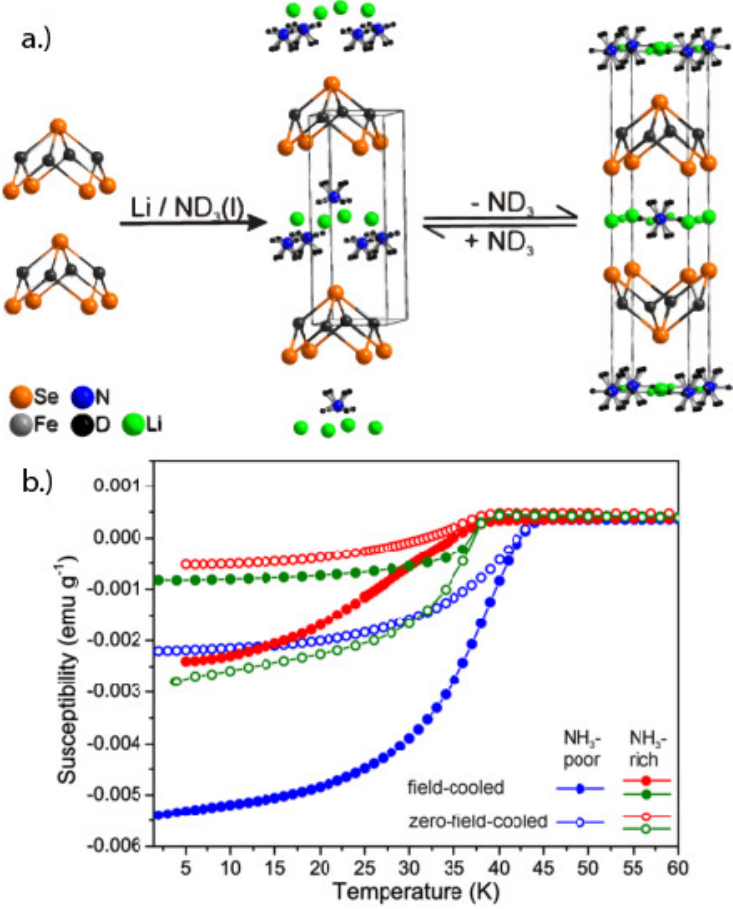}
\caption{a.) The reversible processes by which Li cations and ammonia molecules can intercalate into FeSe. b.) The SQUID magnetometry curves demonstrating that the structure with the double layer of ammonia has a lower $T_c$ than the one with the single layer of NH$_3$, NH$_2^-$, and Li$^+$ species. Reprinted with permission from Ref. \cite{Sedlmaier_2014}. Copyright 2014 American Chemical Society.}
\label{fig_LiNH3-Fe2Se2}
\end{figure}

In a follow up to their neutron diffraction study of Li$_{0.6}$(NH$_2$)$_y$(NH$_3$)$_{1-y}$Fe$_2$Se$_2$, Sedlmaier et al. found that an intermediate phase exists with a double layer of ammonia and amide molecules.\cite{Sedlmaier_2014}  Their \textit{in situ} diffraction experiment revealed that this double layer structure actually retains the primitive tetragonal cell of the parent host, but expanded as shown in Figure \ref{fig_LiNH3-Fe2Se2}.  By ammoniating Li$_{0.6}$(NH$_2$)$_{0.2}^-$(NH$_3$)$_{0.8}$Fe$_2$Se$_2$, Sedlmaier et al. were also able to prepare Li$_{0.6}$(NH$_{2.7}$)$_{1.7}$Fe$_2$Se$_2$, which lowered the $T_c$ from 43(1) K to 37(1) K.\cite{Sedlmaier_2014}   Since this phase was identical to the intermediate phase found in the \textit{in situ} studies, the intercalation of neutral species such as NH$_3$ into these materials is reversible. For similar compounds, Shylin et al. demonstrated that the Li cations do indeed charge dope the FeSe hosts through $^{57}$Fe-M\"ossbauer spectroscopic studies,\cite{Shylin_2015}  therefore confirming that the NH$_2^-$ molecules do not completely charge compensate the Li$^+$ cations.

\begin{figure}[t]
\centering
\includegraphics[width=0.82\linewidth]{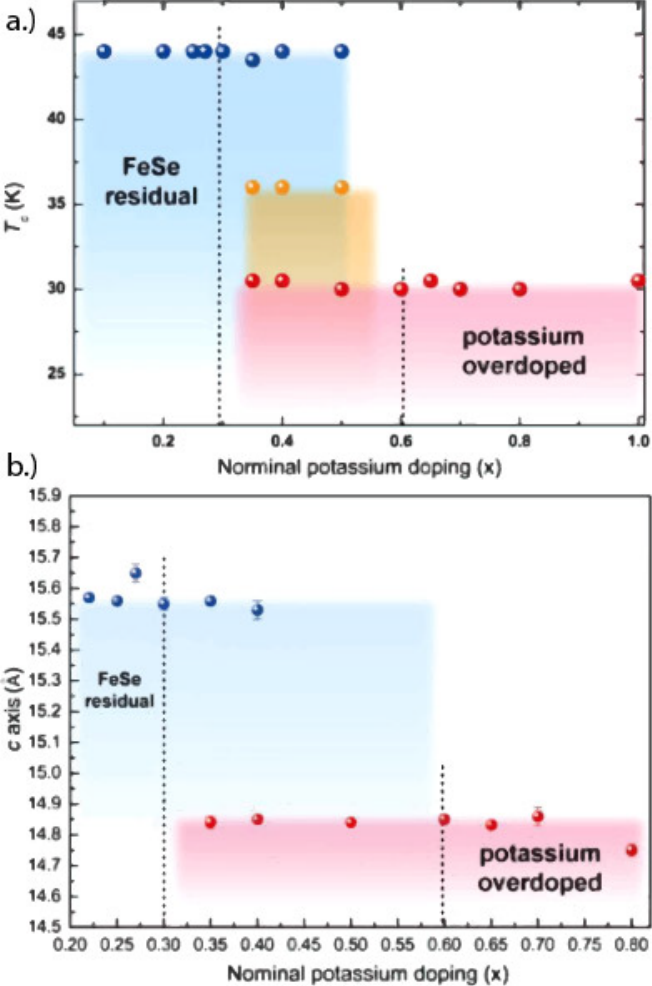}
\caption{The phase diagram for various samples of potassium-intercalated FeSe with the liquid ammonia method.  In a.) the highest-$T_c$ regime exists for samples with the stoichiometry of K$_{0.3}$(NH$_3$)$_y$Fe$_2$Se$_2$, the lowest-$T_c$ is for samples with stoichiometry of K$_{0.6}$(NH$_3$)$_y$Fe$_2$Se$_2$, and the middle regime is an unknown phase.  b.) A clear difference in the $c$ lattice constant is observed, which correlates with $T_c$ and amount of charge doping into the FeSe layers. Reprinted with permission from Ref. \cite{Ying_2013}. Copyright 2013 American Chemical Society.}
\label{fig_KFe2Se2NH3}
\end{figure}

Several groups soon demonstrated that a host of different electropositive metals besides lithium such as potassium, rare earths, and alkaline earths could be intercalated into the FeSe sheets through the liquid ammonia method.\cite{Ying:2012aa, Zheng_2013, Foronda_2015, Zheng_2015, Izumi:2015fy}  Ying et al. found that the amount of K$^+$ cations inserted was more important for affecting $T_c$ than the amount of ammonia intercalated.\cite{Ying_2013}  In their phase diagram, shown in Figure \ref{fig_KFe2Se2NH3}, the superconducting regime does not have a dome-like appearance but rather resembles three flat plateaus.  Essentially, the highest $T_c$ in this system ($\approx 44$ K) is achieved for a stoichiometry of K$_{0.3}$(NH$_3$)$_{0.47}$Fe$_2$Se$_2$ and the lower $T_c$ (30 K), which is coincidentally similar to the one with just alkali species, is found for the composition of K$_{0.6}$(NH$_3$)$_{0.37}$Fe$_2$Se$_2$.   An unidentified phase superconducts below 36 K, which is close to the $T_c$ found by Sedlmaier et al. in their intermediate phase with the primitive cell and a double layer of ammonia molecules.\cite{Sedlmaier_2014}  We can conclude from their study, that there is an optimal amount of doping into the FeSe sheets, which is close to 0.15 electrons per iron cation, and that by overdoping, the $T_c$ is driven down.  It is not clear from this study, however, whether any amide species also intercalates into the FeSe as found for the lithium intercalated compounds by the group of Clarke.\cite{Burrard_2013, Sedlmaier_2014} 

\begin{figure}[t!]
\centering
\includegraphics[width=0.80\linewidth]{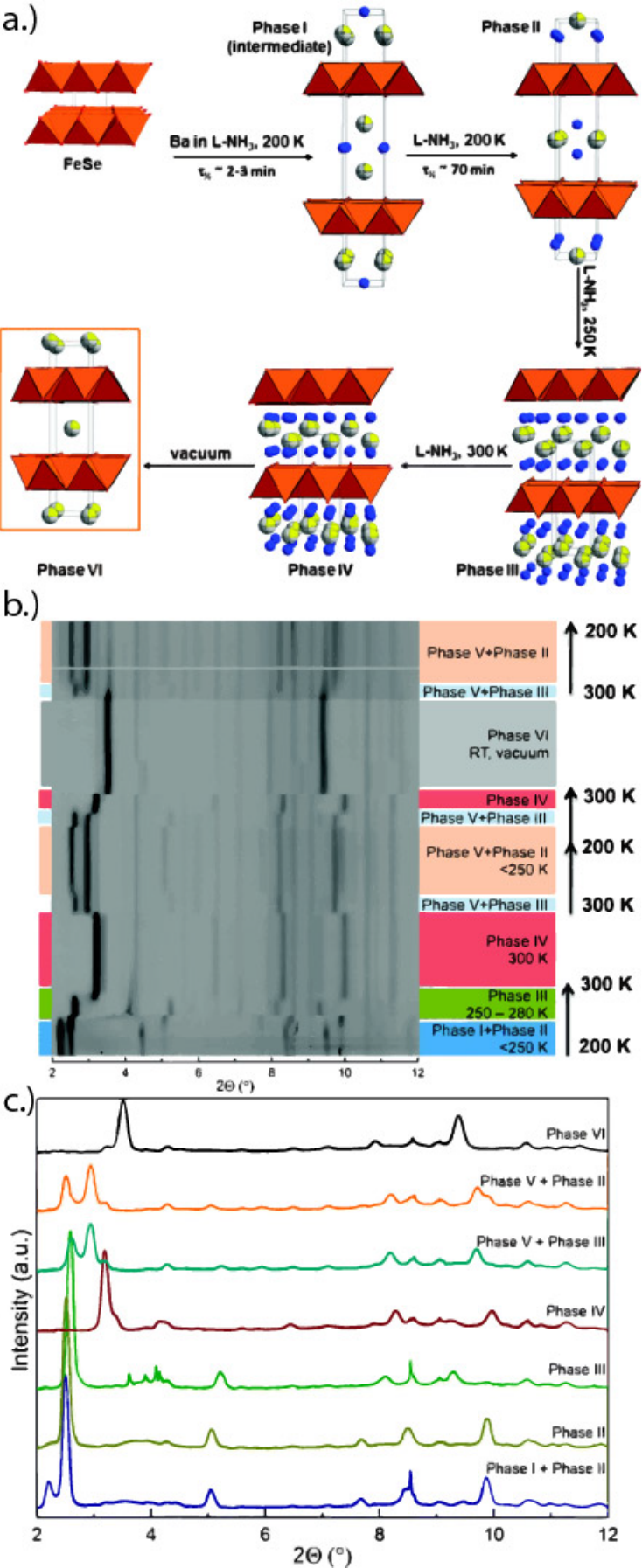}
\caption{a.) The various structures of the barium and ammonia intercalated FeSe phases.  b.) The time resolved X-ray diffraction patterns of the different phases.  c.) The XRD powder patterns corresponding to the various phases represented in a.).  Reprinted with permission from Ref. \cite{Yusenko_2015}. Published by The Royal Society of Chemistry.}
\label{fig_BaNH3-Fe2Se2}
\end{figure}

Through careful \textit{in situ} X-ray diffraction studies, Yusenko et al. found even more intermediate phases during the intercalation of cations via liquid ammonia.\cite{Yusenko_2015}   Initially, a fast intercalation occurs at 200 K with a characteristic time scale, $\tau_{1/2}$, of 2-3 min.  The phase prepared by the fast intercalation includes two layers of Ba$^{2+}$ cations with ammonia molecules in between.  The slower process is the rearrangement of the guest species, $\tau_{1/2}$ = 70 min, to form a phase similar to the one found by Burrard-Lucas et al. for Li,\cite{Burrard_2013} but with more ammonia positions.  Upon warming, the body-centered phases are not stable and decompose to the primitive setting, and upon applying vacuum all the NH$_3$ can be removed to reach a final composition of Ba$_{0.28}$Fe$_2$Se$_2$, which exhibited a $T_c$ of 34 K.  Overall, the highest $T_c$ found was close to 39 K for the ammoniated body-centered phase Ba$_{0.28}$(NH$_3$)$_{1.92}$Fe$_2$Se$_2$.  The entire sequence of powder patterns and the corresponding intercalated phases can be found in Figure \ref{fig_BaNH3-Fe2Se2}. 

\begin{figure}[t!]
\centering
\includegraphics[width=0.95\linewidth]{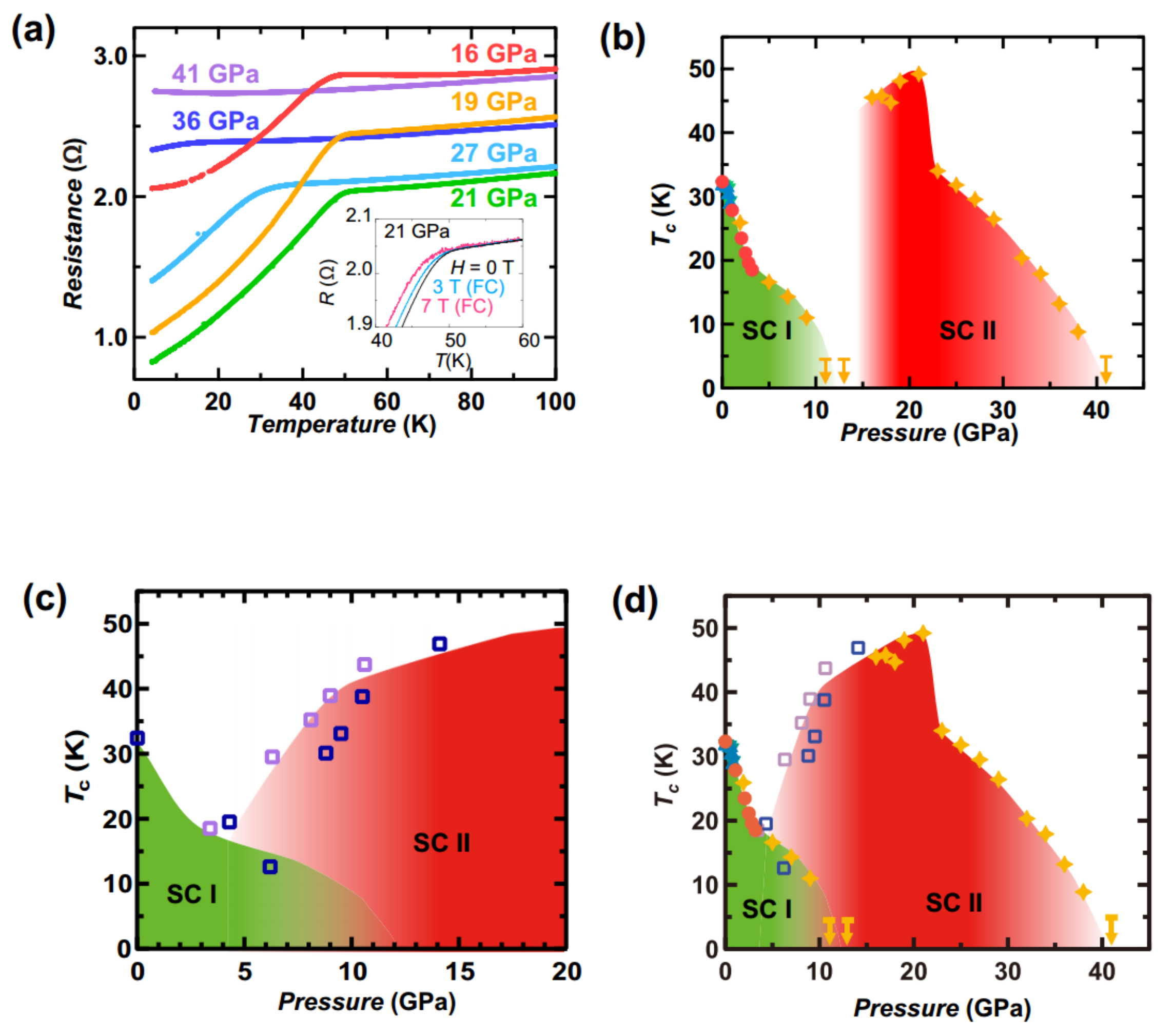}
\caption{a.) Magnetization as a function of pressure for various Cs-intercalated iron selenides  b-d.) Double dome phase diagrams for the Cs-intercalated phases with ammonia.  Reprinted by permission from Ref. \cite{Izumi:2015fy}, Macmillan Publishers Ltd: Nature Communication, copyright 2014.}
\label{fig_CsFeSe}
\end{figure}

Just as FeSe can have its $T_c$ enhanced by the application of pressure, Izumi et al. demonstrated that pressure can also enhance the $T_c$ of the intercalated selenides.\cite{Izumi:2015fy}  After preparing several samples with doped alkali metals, Izumi found a correlation between the tetragonality ($c/a$ ratio) of the unit cell and the $T_c$.   In general, the higher the tetragonality the higher the $T_c$, although the affect seems to plateau beyond a certain $c/a$ ratio.  For their Cs-NH$_3$ intercalated samples, the application of pressure initially drives down the $T_c$, eventually wiping out all superconductivity.  Above the threshold of 14 GPa, however, superconductivity re-emerges and $T_c$ reaches to a high of 49 K until it drops to 0 above 21 GPa.  The resulting phase diagram has a double dome appearance for the superconducting regime as shown in Figure \ref{fig_CsFeSe}, and Izumi et al. ascribe this phenomena to the proximity of a quantum critical point.\cite{Izumi:2015fy}  The authors do not observe any crystallographic phase transitions upon applying pressure. 

\begin{figure}[t!]
\centering
\includegraphics[width=0.90\linewidth]{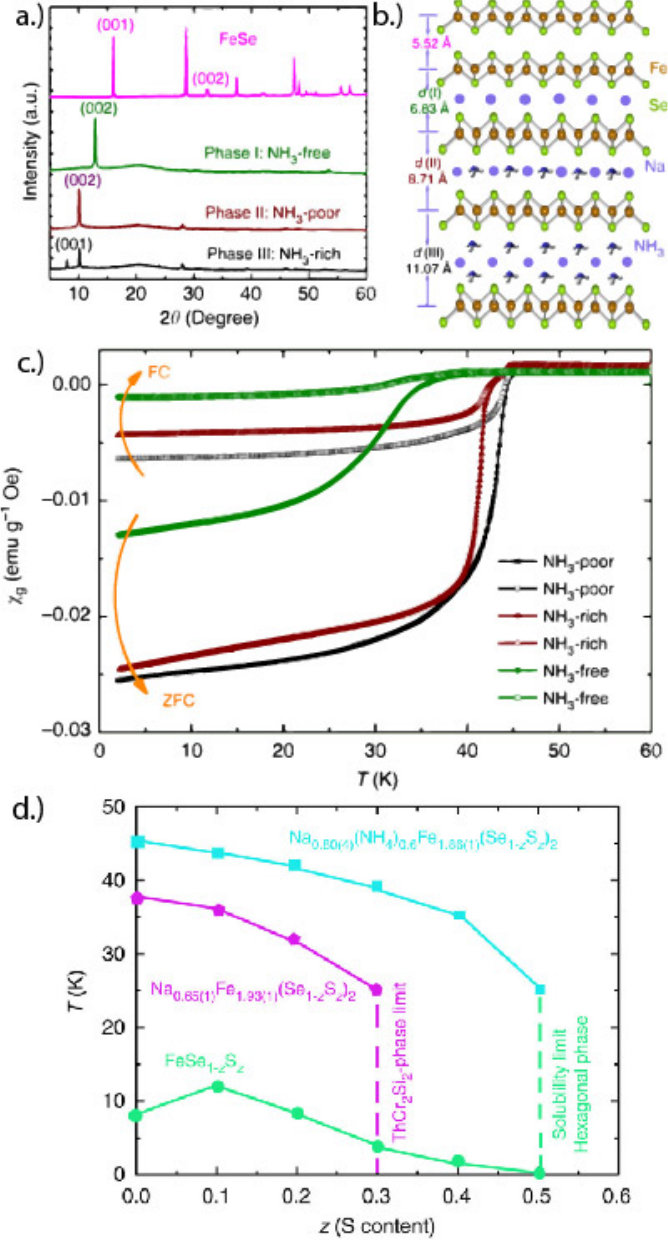}
\caption{a.) The XRD powder patterns of the FeSe and and Na-intercalated phases with either NH$_3$-rich or -poor structures.  b.) A schematic of the structures for the XRD patterns of a.), although the orientation of the ammonia molecule was not solved from the diffraction data.  c.) The SQUID magnetometry data for the Na-intercalated FeSe, and d.) the phase diagram for the Na-Intercalated FeSe$_{1-z}$S$z$ hosts.  Reprinted by permission from Ref. \cite{Guo_2014}, Macmillan Publishers Ltd: Nature Communication, copyright 2014.}
\label{fig_NaFe2SeS}
\end{figure}

\begin{figure*}[h!]
\centering
\includegraphics[width=0.8\textwidth]{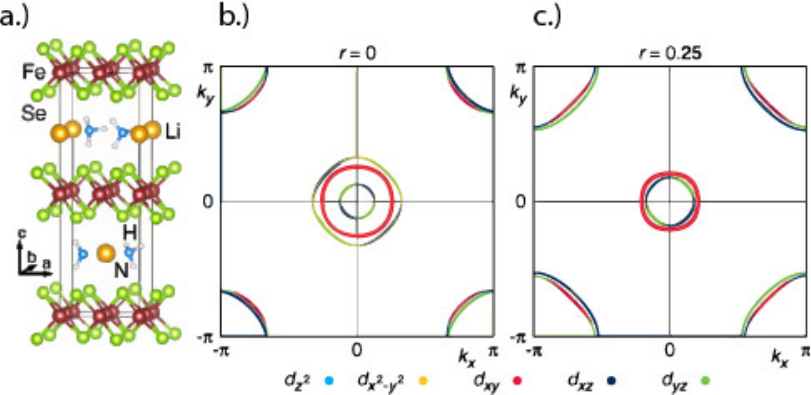}
\caption{a.) The crystal model for the electronic structure calculations of Li$_{0.5}$(NH$_2$)$_{0.5-2r}$(NH$_3$)$_{0.5+2r}$Fe$_2$Se$_2$.  b.) The Fermi surface of the Li-ammonia-intercalated FeSe host for $r=0$, which corresponds to a completely charge compensated Li(NH$_2$) moieties, which do not charge dope the FeSe host. c.) For $r=0.25$, only ammonia is present with Li cations, which contribute 0.25 $e^-$ per Fe cation.  Upon electron doping, the corner-centered electron pockets grow, and the zone-centered hole pockets shrink.  Reprinted figures with permission from Ref. \cite{Guterding_2015}. Copyright 2015 by the American Physical Society.}
\label{fig_LiNH3-FeSeFermiSurf}
\end{figure*}

While no study has been published on the intercalation of amines into FeS or FeTe, some groups have attempted to intercalate mixed anion systems.  For example, while the highest $T_c$ observed for the optimal stoichiometry of FeSe$_{0.5}$Te$_{0.5}$ is 14 K, Sakai et al. demonstrated through intercalation of Li, Na, and Ca cations via the liquid ammonia technique that the $T_c$ could be raised to 26 K, 22 K, and 17 K, respectively.\cite{Sakai_2014, Lei_2014}.   Similarly, Guo et al. intercalated FeSe$_{1-z}$S$_z$ with Na cations and found that the $T_c$ could also be enhanced from that of the of host structure ($T_c \approx 12$ K).\cite{Guo_2014}  As more sulfide is introduce into the lattice, the $T_c$ is diminished and the phase no longer superconducts past a certain amount of S$^{2-}$ substitution.  The phase diagram for their phases is presented in Figure \ref{fig_NaFe2SeS}.  Similar to the conclusions reached by Sedlmaier et al. on Li-intercalation,\cite{Sedlmaier_2014} Guo et al. found a clear relationship between the structure-type and the maximum $T_c$ (Figure \ref{fig_NaFe2SeS}).\cite{Guo_2014}

To understand the various findings from the experimental work on the intercalated iron selenides, Guterding performed electronic structure calculations to understand the role of the doping cations and ammonia spacers on the superconducting properties.\cite{Guterding_2015}  Focusing specifically on the Li$_{0.5}$(NH$_2$)$_{0.5-2r}$(NH$_3$)$_{0.5+2r}$Fe$_2$Se$_2$ system, they found that the major determining factor on the characteristics of the Fermi surface was the amount of electron doping.  As shown in Figure \ref{fig_LiNH3-FeSeFermiSurf}, when the amount of amide present is zero ($r = 0.25$), the Li cations donate approximately 0.25 electrons per Fe cation.  The electron pockets at the corners expand and the hole pockets at the center decrease in size (Figure \ref{fig_LiNH3-FeSeFermiSurf}c).  When amide is added so that the charge of the Li is completely compensated, the opposite occurs and a new hole pocket is created Figure \ref{fig_LiNH3-FeSeFermiSurf}b).  What is also evident from these calculations is that the bands are dominated by the $d_{xy}$, $d_{xz}$, and $d_{yz}$ orbitals.  Guterding et al. conclude that the amount of spacer molecules such as ammonia is not the determining factor for $T_c$, but it can aid in making the Fermi surface more 2D; after that has been accomplished, the amount of electron doping will then enhance the superconducting properties.  

To conclude this section, intercalation of metal cations in liquid ammonia in FeSe hosts seems to be the best manner to prepare higher $T_c$ ($\approx 44$ K) phases that are single phase and therefore free of the insulating antiferromagnetic iron selenides.  The role of the ammonia is not just as a solvent that allows the FeSe hosts to absorb the solvated electrons, but also as a spacer molecule that facilitate the intercalation process.  Furthermore, the ammonia can ionize to form amide in solution with Li$^+$ cations.  So far, no one has reported the formation of the conjugate acid NH$_4^+$ in these intercalated compounds.  During intercalation a series of intermediates are formed as shown by the various \textit{in situ} diffraction studies, and they can be identified on the basis of whether they exhibit primitive or body-centered lattice symmetry.  The spacer aids in further enhancing the 2D nature of the Fermi surface, but the electron doping seems to be the major determining factor of the maximum $T_c$ observed.


\section{Metal intercalation with organic amines}

Just as in the transition metal dichalcogenides, the community soon recognized that other amines besides ammonia could be intercalated into the iron chalcogenides.   The first such report was performed by Krzton-Maziopa et al. where they used anhydrous pyridine as the solvent for a variety of alkali metals including Li, Na, K, and Rb.\cite{Krzton_2012}  By reacting with pure $\beta-$FeSe at 40 $^{\circ}$C, Krzton-Maziopa et al. managed to expand the $c$-axis and enhance the $T_c$ up to 45 K, close to the observed values for the ammonia intercalated selenides.  As shown by powder XRD, the FeSe sheets are indeed intercalated, but the lattice constants however are not that different from those in the ammonia-intercalated samples.  For example, the Li-pyridine intercalated sample had a $c$ parameter of 16.0549 \AA,\cite{Krzton_2012} which is close to the values of 16.1795(6) \AA \, and16.4820(9) \AA \,observed by Burrard-Lucas et al. for the Li-NH$_3$ intercalated samples.\cite{Burrard_2013}  Furthermore, the K-intercalated samples also had a smaller $c$-parameter, which would be unexpected for the larger pyridine ring intercalated with the larger potassium cation.  Krzton-Maziopa et al. also observed LiCN as a byproduct from the reaction, which could indicate that the pyridine rings decomposed and therefore the nature of the amine intercalate could be something other than pyridine.\cite{Krzton_2012} 

\begin{figure}[t!]
\centering
\includegraphics[width=0.90\linewidth]{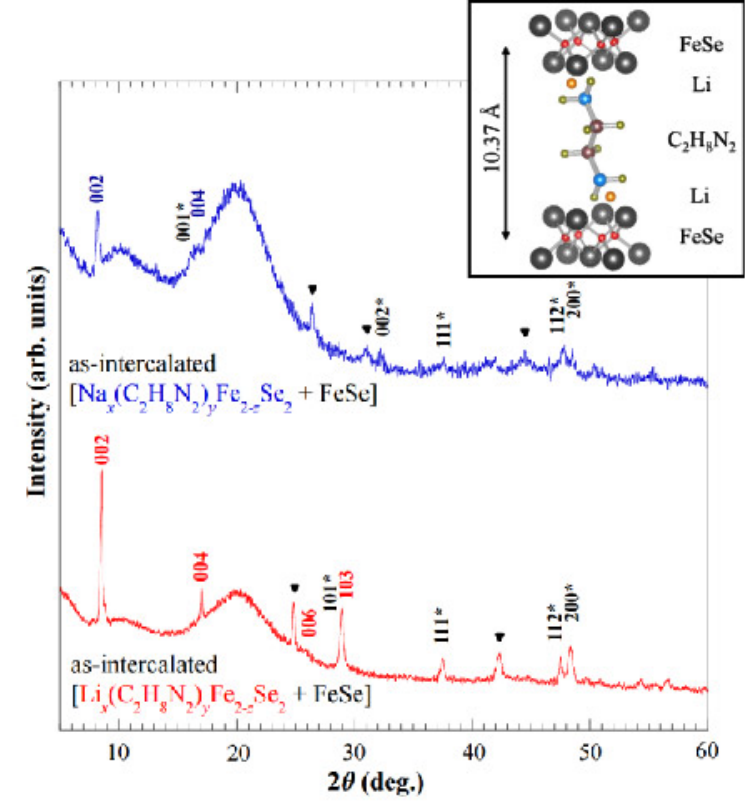}
\caption{Main: The powder XRD patterns of Na- and Li-intercalated FeSe with ethylenediamine acting as both the host and intercalate.  The low angle (002) reflection roughly corresponds to an interplanar spacing close to 10.4 \AA.  Reprinted with permission from Ref \cite{Hatakeda_2014}. A model of the Li-ethylenediamine intercalate in FeSe.  Licensed by IOP Publishing Ltd.  Inset: Reprinted figure with permission from Ref. \cite{Hrovat_2015}. Copyright 2015 by the American Physical Society.}
\label{fig_AEDA-FeSe}
\end{figure}

The work by the group of Noji et al. showed that linear diamines could also be utilized as solvents and intercalates for FeSe.\cite{Hatakeda_2013, Hatakeda_2014, Hosono_2014, Noji_2014, Hrovat_2015}  Starting with the classic coordination ligand, ethylenediamine (C$_2$H$_8$N$_2$), Hatakeda et al. intercalated this ligand along with Li to raise $T_c$ up to 45 K.  The intercalation reaction seems to be sluggish with a total of 7 days at 45 $^{\circ}$C followed by post-annealing at 150 $^{\circ}$C and 200 $^{\circ}$C for several more days under vacuum.\cite{Hatakeda_2013}  Assuming that the compound crystallized in the body-centered cell, the authors indexed the lowest angle peak as the (002) reflection and therefore the interplanar distance to be close to 10.37 \AA \, (Figure \ref{fig_AEDA-FeSe}).  By expanding to even larger chains such as hexamethylenediamine (C$_6$H$_{16}$N$_2$), Hosono et al. increased this interplanar distances to a record 16.225(5) \AA.\cite{Hosono_2014}  This hyper-expansion of the FeSe host, however, does not seem to have increased $T_c$ but rather lowered it to 38 K, which is in line from the previous studies of metal-ammonia intercalation whereby charge doping is more important than the size of the spacer. 

In addition to lithium, other metals can be intercalated with ethylenediamine such as sodium.  Figure \ref{fig_AEDA-FeSe} shows the XRD powder patterns of the Li and Na intercalated compounds along with the C$_2$H$_8$N$_2$ spacers that lead to a body-centered cell and an interplanar distance close to 10.37 \AA \, regardless of the choice of alkali metal.\cite{Noji_2014}  Although the authors claim a $T_c^{onset}$ as high as 57 K from electrical resistivity measurements, the magnetic susceptibility data shows that the true $T_c$ is closer to 45 K, which is more in line with other intercalated FeSe superconductors.\cite{Hatakeda_2014}

A systematic study of various linear diamines with either Na$^+$ or Sr$^{2+}$ cations was carried out by Hayashi et al. to understand the role of charge and spacer length.\cite{Hayashi_2015}  The linear diamines include (H$_2$N)-C$_n$H$_{2n}$(NH$_2$) where $n=$ 0, 2, 3, and 6, and Hayashi et al. found that the interplanar distance could indeed be systematically varied between 8.95 \AA \, and 11.16 \AA \, for the sodium series of compounds.  No correlation, however, was found between the length of the diamine chain and $T_c$.  Instead, the correlation was strictly between charge doping and $T_c$.  For materials with the sodium content varying between 0.80 and 0.93 (per Fe$_2$Se$_2$ unit), the $T_c$ was between 41 K and 46 K, whereas for the strontium content, the $T_c$ was between 34 K and 38 K.  Typically the amount of Sr varied between 0.23 to 0.28 (per Fe$_2$Se$_2$ unit).\cite{Hayashi_2015}  

\begin{figure}[t!]
\centering
\includegraphics[width=1.00\linewidth]{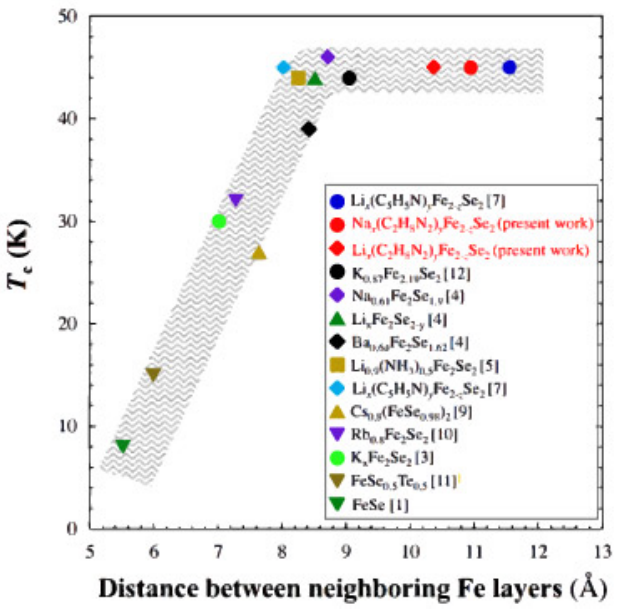}
\caption{The correlation between $T_c$ and the interlayer spacing, $d$, by the intercalation of various species including alkali metal cations, ammonia, and organic amines.  Reprinted from Ref. \cite{Noji_2014}, Copyright 2014, with permission from Elsevier.}
\label{fig_Tcs-FeSe}
\end{figure}

Overall, the recent work of intercalating organic amines with alkali and alkaline earth metals has demonstrated that the maximum $T_c$ is not enhanced beyond that of ammonia with alkali and alkaline earth metals.  The spacer, however does seem to have an effect below a threshold value for the interplanar distance.  As shown in Figure \ref{fig_Tcs-FeSe} for a variety of compounds from FeSe to those with large interlayer spacers, the $T_c$ plateaus above a distance of approximately 8 \AA.  It would be interesting for future work to understand any advantages in utilizing organic amines over ammonia such as thermal stability or finer control over electron doping. However, caution must be exercised since working ethylenediamine can cause fragmentation of the FeSe layers such as Fe$_3$Se$_4$(en)$_2$.\cite{Kovnir_2013}

\begin{figure}[t]
\centering
\includegraphics[width=0.90\linewidth]{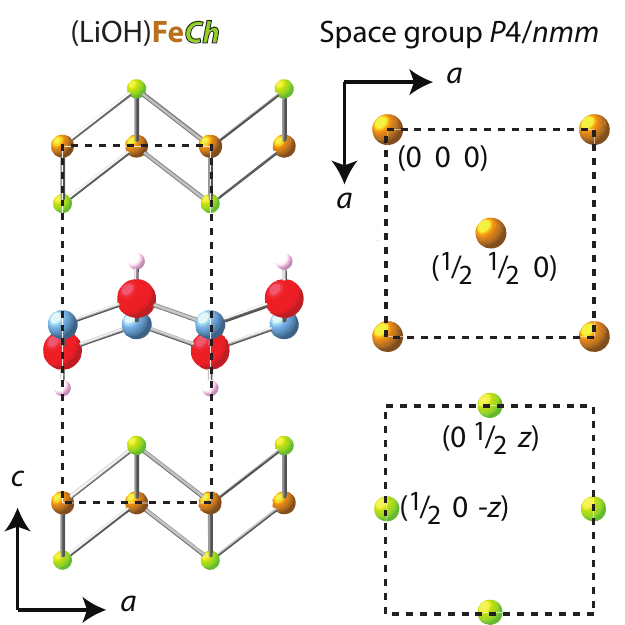}
\caption{The crystal structure of (LiOH)FeSe, which has a similar structure to the ZrCuSiAs-type structure.  In the superconducting phases, Fe$^2+$ cations are doped into the Li sites, which charge dopes the FeSe hosts.  The LiOH intercalate has the anti-PbO-type structure.}
\label{fig_LiOHFeSe}
\end{figure}

An interesting development in the intercalation chemistry of FeSe was the demonstration by Lu et al. that such reactions could occur in aqueous solutions as opposed to ammoniacal ones that were strictly anhydrous.\cite{Lu_2014}  Under hydrothermal conditions, Lu et al. added a selenide source in the form of selenourea to a strongly basic solution of LiOH along with iron metal.  The resulting compound was formulated as LiFeO$_2$Fe$_2$Se$_2$, which was structurally related to the LnOFeAs superconductors and the ZrCuSiAs-type structure (Figure \ref{fig_LiOHFeSe}).  An electronic structure calculations study by Heil et al. predicted that the ground state of this oxide-intercalated FeSe should be that of an antiferromagnet due to the effect of the $d^5$ states (from Fe$^{3+}$) in the oxide layer.\cite{Heil_2014} However, this work did not yet have the full experimental details of the chemical composition and missed the protons present in the oxide layer, which would therefore call into question whether Fe$^{3+}$ is really present in this layer as opposed to Fe$^{2+}$.

Soon after the work of Lu et al., Pachmayr et al. were able to prepare small single crystals from hydrothermal methods (Figure \ref{fig_LiOHFeSe-SEM}) and found residual electron density corresponding to hydrogen from their diffraction measurements.\cite{Pachmayr_2015b}  Therefore, Pachmayr et al. formulated the intercalate to actually be a hydroxide instead of oxide, and their single crystal X-ray diffraction results showed that the composition is close to (Li$_{0.8}$Fe$_{0.2}$OH)(Fe$_{0.915}$Li$_{0.085}$Se).  The hydroxide intercalate has a structure very similar to LiOH, which itself crystallizes in the anti-PbO-type structure.  Furthermore, antisite substitution occurs so that some iron cations substitute onto the Li site, and \textit{vice versa}.  Pachmayr et al. also established that the nature of the iron in the hydroxide layer is Fe$^{2+}$ and not Fe$^{3+}$ based on the Fe--O distance of 201.6 pm.  The  Fe$^{2+}$  cations act to both electron dope into the FeSe host and to flatten the (Li/Fe)O$_4$ tetrahedra since Fe$^{2+}$ prefers more square planar coordination.\cite{Pachmayr_2015b}

\section{Metal hydroxide intercalation under hydrothermal conditions}

\begin{figure}[t!]
\centering
\includegraphics[width=0.90\linewidth]{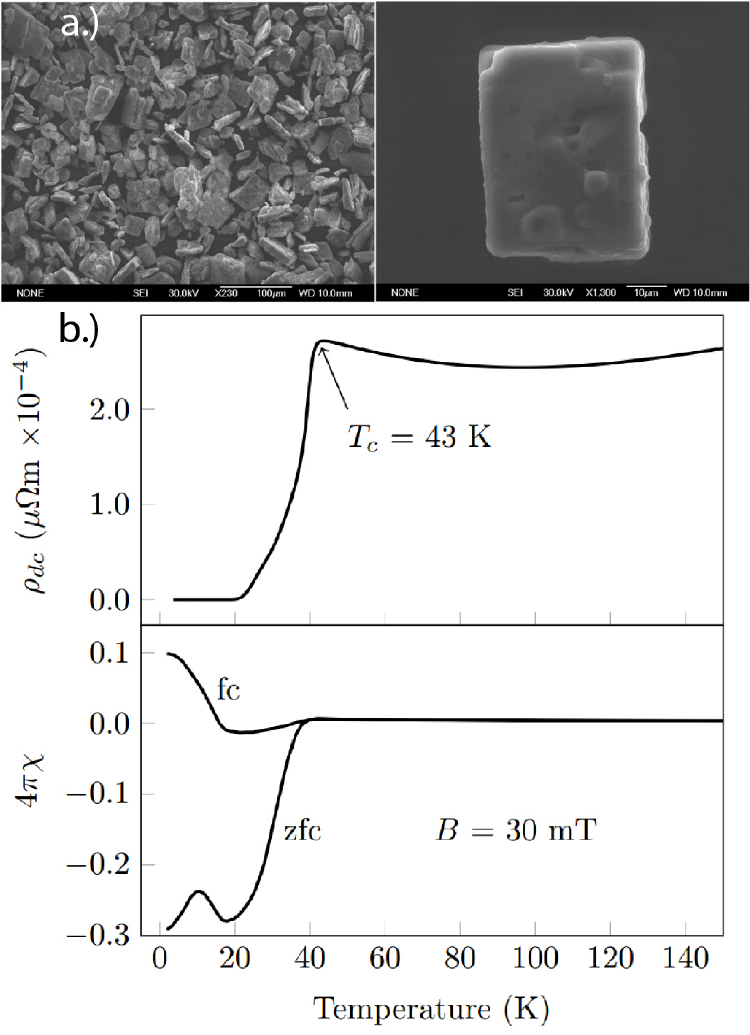}
\caption{a.) The SEM image of the (Li$_{0.8}$Fe$_{0.2}$OH)Fe$_{0.915}$Li$_{0.085}$Se single crystals prepared by hydrothermal synthesis.  b.) The electrical resistivity and magnetic susceptibility of (Li$_{0.8}$Fe$_{0.2}$OH)Fe$_{0.915}$Li$_{0.085}$Se, and the low temperature signal below 20 K represents a ferromagnetic response that coexists with superconductivity. Reprinted with permission from Ref. \cite{Pachmayr_2015b}. Copyright 2015 Wiley-VCH Verlag GmbH and Co. KGaA, Weinheim.}
\label{fig_LiOHFeSe-SEM}
\end{figure}

Interestingly, due to the iron substitution onto the hydroxide layer, Pachmayr et al. found the compound to express both ferromagnetism and superconductivity at base temperature, which led to an interesting phenomena by which spontaneous vortex lattice forms in this type-II superconductor.\cite{Pachmayr_2015b}  As shown in Figure \ref{fig_LiOHFeSe-SEM}, the zero-field cooled curve in the magnetic susceptibility shows and a ferromagnetic signal even in the diamagnetic regime. Through nuclear magnetic resonance (NMR) studies of this compound, Lu et al. found the hydroxide layer to actually exhibit antiferromagnetism instead of ferromagnetism.\cite{Lu:2015aa}  A small angle neutron scattering (SANS) study found that under a small applied magnetic field, a vortex lattice was found in (Li$_{1-x}$Fe$_x$OD)FeSe in the pattern with a characteristic length not unlike in other superconductors.\cite{Lynn_2015}  Lynn et al. concluded, however, that the vortex forms only under an applied magnetic field and is not spontaneous.

\begin{figure}[t]
\centering
\includegraphics[width=0.90\linewidth]{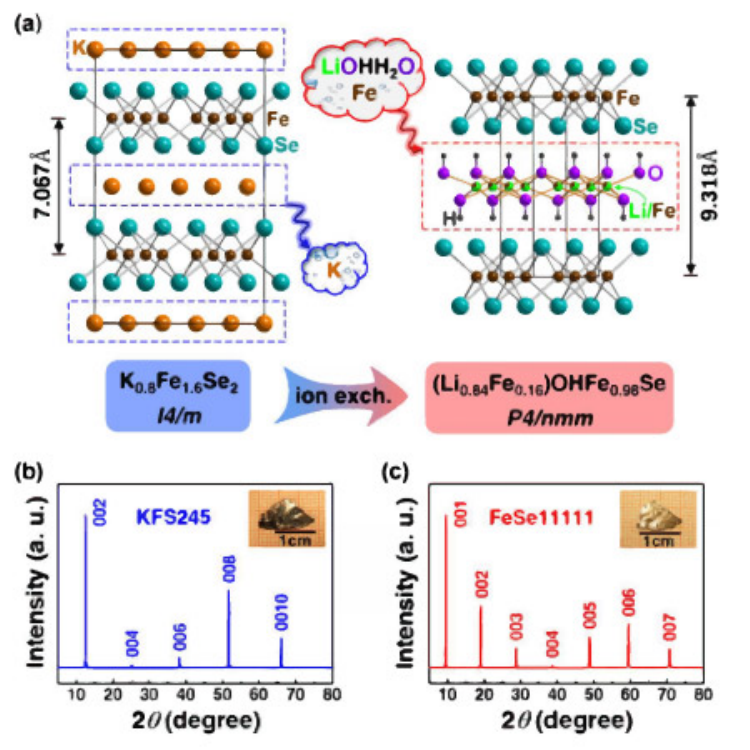}
\caption{A schematic of the cation exchange reaction under hydrothermal and strongly basic conditions to convert K$_x$Fe$_{2-y}$Se$_2$ to (LiOH)FeSe.  b.) The resulting change in the powder patterns from a body centered structure to the primitive cell.  Reprinted figure with permission from Ref. \cite{Dong_2015}. Copyright 2015 by the American Physical Society.}
\label{fig_ionex-FeSe}
\end{figure}

Several studies on the phase diagram of the (Li$_{1-x}$Fe$_x$OH)FeSe then followed to understand how the structural and chemical parameters controlled $T_c$.  Sun et al. found that instead of preparing the hydroxide samples \textit{in situ} under hydrothermal conditions, that FeSe could act as the source for the preparation of (Li$_{1-x}$Fe$_x$OH)Fe$_{1-y}$Se where ($x \approx 0.2$ and $0.15 > y > 0.02$).\cite{Sun_2015}  The samples prepared in this manner repeatedly led to lower $T_c$'s, however, so Sun et al. performed post-synthetic modification of these powder samples by treating them with an ammoniacal solution of Li metal.  This reductive post-synthetic modification led to maximal $T_c$ of 45 K, and Sun et al. explain this result as filling in iron vacancies in the FeSe layer.\cite{Sun_2015}  Indeed, the technique of post-synthetic modification is a form of intercalation chemistry.  The large spread in iron deficiency found in their samples led Sun et al. to quantitatively correlate iron occupancy in the selenide layer to superconductivity and to conclude that full iron occupancy in the selenide layer along with iron with an oxidation state below +2 are required for superconductivity. 

The area of soft chemistry can be further expanded beyond powder to include the preparation of single crystal samples of (Li$_{1-x}$Fe$_x$OH)FeSe.  Dong et al. found that starting with K$_x$Fe$_2$Se$_2$ single crystals prepared at high temperatures, one could perform a cation exchange under hydrothermal conditions to prepare LiOHFeSe single crystals.\cite{Dong_2015}  Such exchange and the resulting X-ray diffraction patterns are shown in Figure \ref{fig_ionex-FeSe}  Both the work of Dong et al. and Sun et al. demonstrate that the iron selenides exhibit considerable soft chemistry that allows for facile preparation of superconductors with optimal properties.

Zhou et al. were able to construct two phase diagrams, one for (LiOH)FeSe and the deuterated (LiOD)FeSe from single crystal and powder samples.  No isotope effect was observed from the deuterium substitution.  In their phase diagrams, Zhou et al. correlate the $T_c$ to the tetragonality of the unit cell ($c/a$ ratio), which might be related to the overall oxidation state of the iron in the FeSe layer and therefore to the amount of charge doping from the hydroxide layer.\cite{Zhou_2016}  Dong et al. also correlated $T_c$ to the lattice parameters but added an antiferromagnetic phase for samples with very small $c$ lattice constant, which they referred to as the parent compound to the superconductor.\cite{Dong_2015a}  This antiferromagnetic ordering observed by Dong et al. in their magnetization studies, however, was disputed by Zhou et al. since they demonstrated that under hydrothermal conditions various ferrimagnetic iron oxides such as magnetite can be prepared.  Furthermore, through their neutron diffraction studies of deuterated samples, Zhou et al. found no evidence for long range antiferromagnetic ordering either in superconducting or non-superconducting samples.\cite{Zhou_2016}  This result from Zhou et al. would be in line with the vast literature on iron selenides, where a parent compound with long-range antiferromagnetic ordering has yet to be found.\cite{Ivanovskii_2011}  

The insertion of hydroxide layers into other hosts has been accomplished as well, notably mackinawite FeS.  While the host FeS has very recently been shown to be superconducting with a $T_c$ close to 5 K, Pachmayr found the LiOH-intercalated phase of FeS to be non-superconducting and instead displays only weak ferromagnetism.\cite{Pachmayr_2015}  Likewise, Lu et al. found that in hosts with mixed anions such as FeSe$_{1-z}$S$_z$, lithium hydroxide intercalation could be achieved but that the $T_c$ drops off dramatically as the amount of sulfide is introduced into the lattice.\cite{Lu_2014_2}  This could be due to the general instability of mackinawite FeS, which might cause sufficient iron vacancies to form upon intercalation so that they completely disrupt the conducting properties.

\section{Conclusions and future directions}

The iron chalcogenides, and in particular FeSe, have been proven to be versatile hosts for intercalation chemistry.  Furthermore, intercalation chemistry affords superconducting samples that are phase pure and therefore reveal the true ground state properties.  These low-temperature synthetic techniques are superior to that of simply melting the constituent elements (alkali metal, iron, and selenium), since the latter leads to phase separation and the inclusion of an antiferromagnetic, insulating phase.  The intercalation chemistry of FeSe therefore demonstrates that solid state chemists will continue to play a role in furthering our understanding of superconducting materials.

The major parameters that seem to be important across the entire series of iron chalcogenides include structure and charge doping.  As such, we have gathered the relevant parameters across various intercalated selenides in Table \ref{parameters} for comparison.  Although there have been many efforts to correlate $T_c$ with charge doping there are many inconsistencies apparent in Table \ref{parameters}, and this may be on account of the difficulty on determining whether the amines intercalated are neutral or charged.  Furthermore, due to the nature of the disordered intercalates and sometimes low-quality of the powder patterns, more detailed crystallographic data such as anion height and $Ch$--Fe--$Ch$ bond angles are difficult to gather and compare presently.  Therefore, more detailed and systematic studies correlating these key parameters should be carried out in future studies.  The consensus so far is that the intercalates should separate the FeSe layers sufficiently in order to enhance the two-dimensionality of the electronic structure and also charge dope it enough to enhance the coupling between the electron and hole pockets of the Fermi surface.  The data from the studies reviewed here suggest that the upper limit for bulk intercalated-FeSe is 45 K.

\begin{table}[b!]
\caption{Structural parameters for FeSe and various intercalated FeSe superconductors.  Here $d$ refers to the interplanar spacing, or the spacing between the Fe sublattices, as determined by diffraction experiments at room temperature.  In compounds with the primitive cell, this is the (001) reflection, and in those with the body-centered cell, the (002) reflection.}
\resizebox{\columnwidth}{!}{
\begin{tabular}{l l l l l}
\hline
compound															&  $d$ (\AA)		& 	$T_c$ (K)		&  $e^-$/Fe cation	& Reference \\
\hline
\hline
FeSe																						&  5.5258(1)			&  	8					& 0.00	& \cite{McQueen_2009} \\
K$_{0.38(2)}$Fe$_{2.1(3)}$Se$_2$											&	7.1329(1)			&	28					& 0.18	& \cite{Shoemaker_2012} \\
Li$_{0.6}$(ND$_2$)$_{0.2}$(ND$_3$)$_{0.8}$Fe$_2$Se$_2$		&  8.2410(5)			&  43(1)				& 0.20  	& \cite{Burrard_2013}\\
Li$_{0.6}$(ND$_2$)$_{0.34}$(ND$_3$)$_{1.36}$Fe$_2$Se$_2$*	&  10.30704(8)		&  39(1)				& 0.13  	& \cite{Sedlmaier_2014}\\
K$_{0.3}$(NH$_3$)$_{0.47}$Fe$_2$Se$_2$								&	7.78(1)				&	44					& 0.15	& \cite{Ying_2013} \\
K$_{0.6}$(NH$_3$)$_{0.37}$Fe$_2$Se$_2$								&  7.42(1)				&  31					& 0.30	& \cite{Ying_2013} \\
Ba$_{0.37}$(NH$_3$)$_{1.04}$Fe$_2$Se$_2$							&	9.060					&	36					& 0.37	& \cite{Yusenko_2015} \\
Na$_{1.0}$(C$_2$H$_8$N$_2$)$_{1.5}$Fe$_2$Se$_2$				&	9.51					&	46.5				& 0.5		& \cite{Hayashi_2015} \\
Sr$_{0.5}$(N$_2$H$_4$)$_{3.0}$Fe$_2$Se$_2$						&	8.91					&	35					& 0.5		& \cite{Hayashi_2015} \\
(Li$_{0.837}$Fe$_{0.165}$OH)FeSe											&	9.3512(2)			&	40					& 0.165	& \cite{Sun_2015} \\
(Li$_{0.828(1)}$Fe$_{0.172(1)}$OH)FeSe*									&	9.1330(2)			&	37					& 0.172 & \cite{Zhou_2016} \\
\hline
\multicolumn{5}{l} {* diffraction data taken between 5 K to 7 K} \\
\end{tabular}}
\label{parameters}
\end{table}

From a historical perspective, it is interesting to compare the intercalated iron chalcogenides with that of other 2D hosts such as the transition metal dichalcogenides (see section \ref{MCh2}) and the Na$_y$CoO$_2$ hydrates.  There seems to be sufficient evidence across these 2D materials that hydrogen bonding has a large role to play in their stability.  The ammonia molecules in both FeSe and the dichalcogenides such a TaS$_2$ are oriented so that the N--H bonds and not the lone pair of electrons on N point towards the chalcogenide anions. In a neutron diffraction study of NH$_3$ intercalated TaS$_2$, Young et al. found evidence for hydrogen bonding in the form of N--H$\cdot \cdot \cdot \cdot$S bonds where the S--H interatomic distances ranged between 2.49 \AA \, and 2.66 \AA.\cite{Young_1990}  Similarly, Burrard-Lucas et al. found evidence for N--H$\cdot \cdot \cdot \cdot$Se bridges in the Li-NH$_3$ intercalated FeSe.\cite{Burrard_2013}  An interesting difference between the iron chalcogenides and the transition metal dichalcogenides is that in the iron chalcogenides the ammonia does not enter the host structure as the conjugate acid (NH$_4^+$), but rather as the conjugate base NH$_2^-$ along with neutral ammonia molecules.  Favoring amide formation rather than ammonium makes sense in light of the fact that the FeSe layers are anionic and therefore have electron density at the Fermi level to reduce ammonia.

Similar to the cobaltates (see section \ref{cobaltates}), the iron chalcogenides also have significant aqueous chemistry, and the ability to integrate hydroxides such a LiOH.  The weak van der Waals forces present in the chalcogenides are presumably replaced by ionic forces from the cationic intercalates.  However, there might be more than these ionic forces participating in interlayer bonding.   Given the hygroscopic nature of Na$_y$CoO$_2$ and the orientation of the hydroxyl groups in (Li$_{1-x}$Fe$_x$OH)FeSe, hydrogen bonding should also play a role in the phase stability of these materials.  Indeed, neutron diffraction study by Zhou et al. (Li$_{1-x}$Fe$_x$OD)FeSe found crystallographic evidence for O--D$\cdot \cdot \cdot \cdot$Se bonding where the Se--D distance was approximately 3.03 \AA.\cite{Zhou_2016}

The intercalation chemistry of FeSe also has similarities with the superconducting fullerenes (see section \ref{fullerenes}).  The amount of charge doping in both sets of materials seems to be critical to raising the $T_c$.  In the case of intercalated fullerenes, the alkali metals such as potassium electron dope the conduction band and since this band is narrow, the density of states at the Fermi level is high.  Similarly, FeSe requires a certain amount of charge doping at the Fermi level to enhance its superconducting properties.  Unlike the doped fullerenes, however, the iron chalcogenides are multi-band superconductors where the Fermi surface has both hole and electron pockets instead of just one conduction band.  The nesting of these multiple bands seems to be critical in the case of FeSe.

Even if the maximum $T_c$ has been achieved in FeSe, there remain multiple future directions for this field.  One, the orientation of the of the organic amines intercalates in FeSe has not been solved nor the nature of the guest species after intercalation.  In other words, are the organic amines neutral or their conjugate bases upon intercalation?  This question is important to establish the correct electron doping for optimal superconducting properties, which could help answer the question as to the mechanism for such a high $T_c$.  The coordination of the organic amines, \textit{e.g.} pyridine and ethylenediamine, could also be solved in order to determine the importance of hydrogen bonding and the intercalated cations in stabilizing these compounds. The work of Kovnir et al. in the intercalation of structurally related iron selenides with intercalated organic amines demonstrates how this problem could be solved from advanced diffraction techniques.\cite{Greenfield_2015}

Another set of opportunities in this field is to explore beyond FeSe.  The intercalation chemistry of mackinawite FeS, for example, has not been fully explored.  Since it has just been found to be superconducting at 5 K, there exists an opportunity to discover higher $T_c$ sulfides.  Finally, other metals besides iron could be investigated such as cobalt, nickel, and copper, which have similar chalcogenide chemistry to iron.  These new intercalated transition metal monochalcogenides could represent a new area of materials chemistry in a similar fashion to the transition metal dichalcogenides.

\section{Acknowledgements}

We thank the National Science Foundation CAREER, DMR-1455118, for financial support.



\end{document}